\documentclass[AMA,STIX1COL]{WileyNJD_v2}
\usepackage{moreverb}
\usepackage[caption=false,font=normalsize,labelfont=sf,textfont=sf]{subfig}
\usepackage{rotating}
\usepackage[export]{adjustbox}
\usepackage{booktabs}
\usepackage{longtable}

\newcommand\BibTeX{{\rmfamily B\kern-.05em \textsc{i\kern-.025em b}\kern-.08em
T\kern-.1667em\lower.7ex\hbox{E}\kern-.125emX}}

\articletype{Article Type}%

\begin{document}

\title{A Compact Quasi-Yagi Antenna for FMCW Radar-on-Chip based Through-Wall Imaging}

\author[1]{Anand Kumar*}

\author[1]{Easha *}

\author[1]{Debdeep Sarkar}

\author[1]{Gaurab Banerjee}

\authormark{Anand Kumar \textsc{et al}}

\address[1]{\orgdiv{Department of Electrical Communication Engineering}, \orgname{Indian Institute of Science, Bengaluru}, \orgaddress{\state{Karnataka}, \country{India}}}




\corres{Anand Kumar and Easha, \newline *The authors have equally contributed to this paper. \newline Department of Electrical Communication Engineering, Indian Institute of Science, Bengaluru, KA 560012, India. \newline \email{anandkumar13@iisc.ac.in, easha1@iisc.ac.in}}

\presentaddress{Department of Electrical Communication Engineering, Indian Institute of Science, Bengaluru, KA 560012, India.}

\abstract[Abstract]{A compact quasi-Yagi antenna with a modified ground plane is designed for a through-wall radar (TWR) on-chip. A slot-based ground plane modification in the proposed antenna results in significant miniaturization with an increase in the impedance bandwidth by 44.62\%. The antenna has a high directivity of 9.02~dBi and a front-to-back ratio of 25.76~dB at 2.4~GHz. Based on experiments in real-world deployment scenarios, the performance of the proposed quasi-Yagi antenna is found to be comparable to that of a Vivaldi antenna and a commercial-off-the-shelf (COTS) horn antenna. Spectrogram-based signatures of a moving person behind a wooden partition and a 40~cm thick masonry wall are successfully obtained using the designed antenna, demonstrating the suitability of the quasi-Yagi antenna for portable applications using a radar-on-chip.}

\keywords{Antenna, Vivaldi, Quasi-Yagi, Directors, Ultra wide-band, UWB, Chirp, Through-wall radar, TWR, Spectrogram, FMCW }


\maketitle

\footnotetext{\textbf{Abbreviations:} UWB, Ultra wide-band; FMCW, frequency modulated continuous wave; TWR, Through-wall radar}

\section{Introduction}
A through-wall radar (TWR) based imaging system enables the determination of target locations behind walls and the remote-mapping of the contents of a room. Such an imaging system is of special interest to the law enforcement, defence, and search and rescue departments in various countries  \cite{fontana2004recent,yang2009development,sensors}. Specific applications include concealed object detection at airports, thru-wall imaging for intelligence gathering at terrorist held buildings and the rescue of people buried under rubble in collapsed buildings \cite{8556006}.
The electromagnetic (EM) waves at the low microwave frequency range can penetrate common building materials, enabling the radar to image behind the wall. However, due to the complexity of the scattering scenario, the radar signal undergoes multipath propagation phenomena. These typically manifest themselves as environmental clutter, which may impair the detection and tracking of actual targets \cite{8708969,yamamoto2021aerial,9250497}.
\begin{figure}
    \centering
    \includegraphics[width=\columnwidth,height=8cm,keepaspectratio]{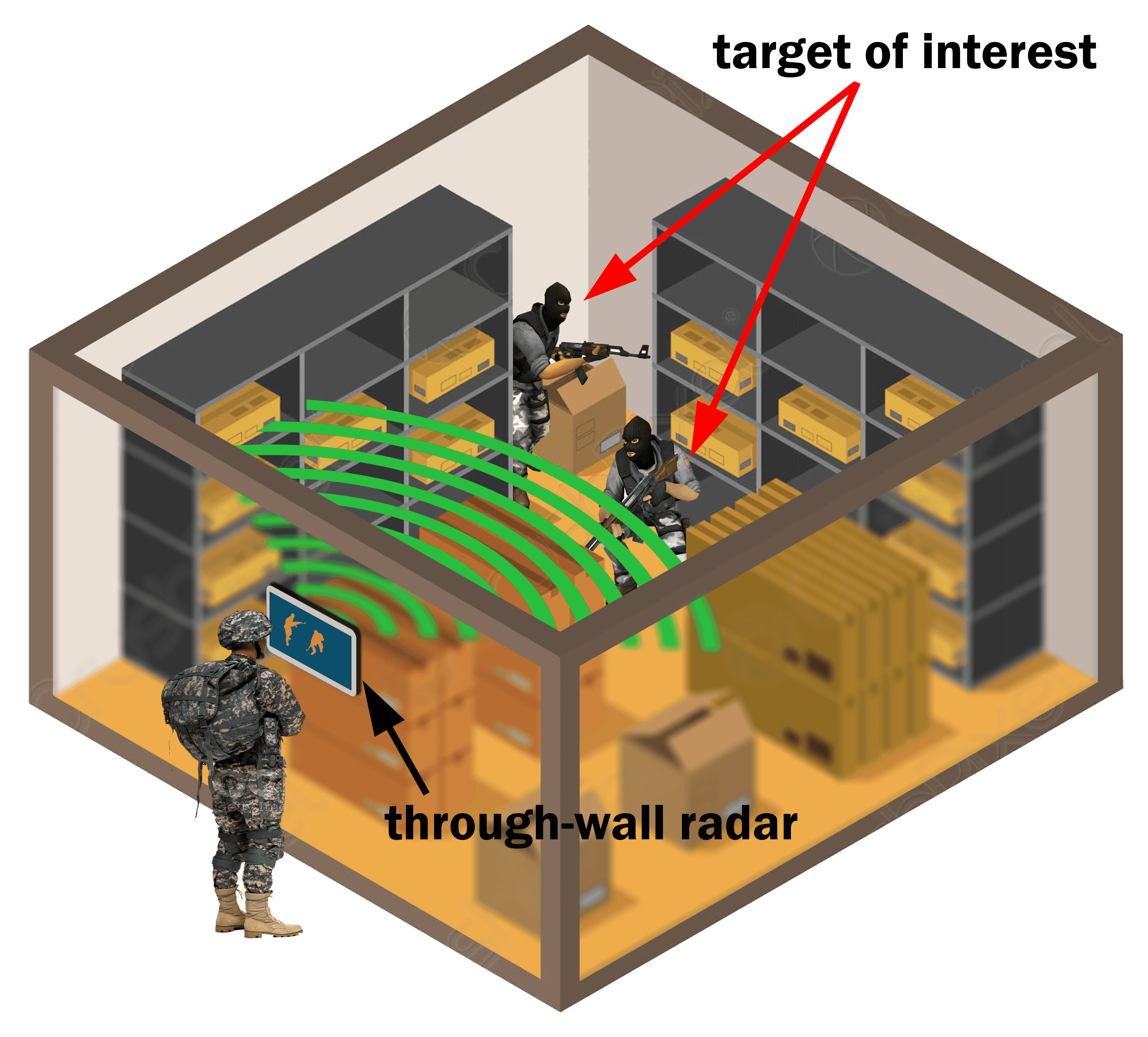}
    \caption{Pictorial representation of a TWRoC in action.}
    \label{radarroom}%
    \vspace{-0.5cm}
\end{figure}

Ultrawideband (UWB) TWR technology has excellent potential due to the larger frequency range over which performance variation due to propagation characteristics of different building materials can be handled better \cite{amin2017through,an2021range}. An improved understanding of the TWR propagation mechanism and high-resolution, time-efficient imaging techniques, can enhance this technology \cite{yang2009development, ahmad2005synthetic,wang2006imaging,song2005through}. 
It is recommended to use ultrawideband (UWB) modulation technology in order to deal with indoor propagation and to achieve improved range resolution for human localization \cite{building}. 
In \cite{snowpack}, L-band frequencies are used to develop a real-aperture FMCW radar system for the detection of objects buried in heavily wet snow-pack. In \cite{mmwave}, a millimeter-wave frequency-modulated continuous wave radar is used for through-wall surveillance applications. However, the wall attenuation increases with the frequency of operation thereby decreasing the penetration capability \cite{amin2017through}.
The development of a robust TWR system requires extensive theoretical and experimental studies involving the entire system, and its components such as the radar subsystem and the antenna.

Through-wall radar systems provide significant challenges to antenna design due to varying levels of attenuation caused by different types of construction material such as concrete, brick and mortar, wood and glass. Hence a TWR antenna should have high directivity and gain. A compact ultrawideband antenna array for TWR applications is proposed in \cite{ren2009compact} and a two-element L-band quasi-Yagi antenna array in \cite{4388128}, but the antenna has an omnidirectional radiation pattern. Portability often places a practical limit on antenna size for a given angular resolution. Therefore the antenna has to satisfy size and weight constraints. Ground-based radar systems, specifically TWR systems, operate in extremely complex propagation environments. It is therefore necessary to perform a comprehensive study of the radar system in different settings to develop a robust system \cite{burchett2006advances}. A patch antenna is optimized for through-wall detection applications in \cite{patch}. However, the form factor of the antenna is larger than that of the compact quasi-Yagi design proposed in this work. The Vivaldi antenna has also been used in TWR systems \cite{yang2009development,hu2019design,patch}. However, a more compact antenna working in the desired band with satisfactory gain and radiation efficiency is required to optimize space in portable applications.

This paper presents a compact planar quasi-Yagi antenna with a modified ground plane. A slot is introduced near the feed of the dipoles of the quasi-Yagi antenna in the ground plane which leads to increased effective electrical length, thereby causing miniaturization. Flaring is introduced in the arms of the dipoles to increase the impedance bandwidth (IBW), and directors are added to further improve the matching in the band. The proposed antenna offers a high gain of 7.8-9.8~dBi in the entire band of interest. Its size and performance are compared with planar quasi-Yagi antennas reported in recent literature.

The rest of this paper is organised as follows. In section \ref{Through wall radar}, the basic working principle of an FMCW radar is described with the specifications of a radar-on-chip used in this study. Section \ref{Antenna Design} discusses the antenna design for the radar. The design process and parameters of a Vivaldi antenna are discussed in section \ref{Vivaldi Antenna}, while the development of the quasi-Yagi antenna and its miniaturisation is described in section \ref{Quasi-Yagi Antenna}. Section \ref{Experimental Results and Analysis} presents experimental data obtained from radar system tests and its analysis in the context of specific deployment scenarios. Section \ref{conclusion} concludes the paper.

 \section{Through wall Radar} \label{Through wall radar} 
Unmodulated continuous wave (CW) radars can accurately measure target radial velocity (using Doppler shift) and angular position. However, due to the continuous nature of emission, range measurement is not possible without modifications to the radar operation and waveforms. A frequency modulated continuous wave (FMCW) radar can measure both range and Doppler information \cite{Yan2022parameter}, which makes it valuable for TWR applications.
\subsection{FMCW Radar}
\begin{figure}[!t]
\centering
\subfloat[]{\includegraphics[width=0.4\columnwidth,valign=b]{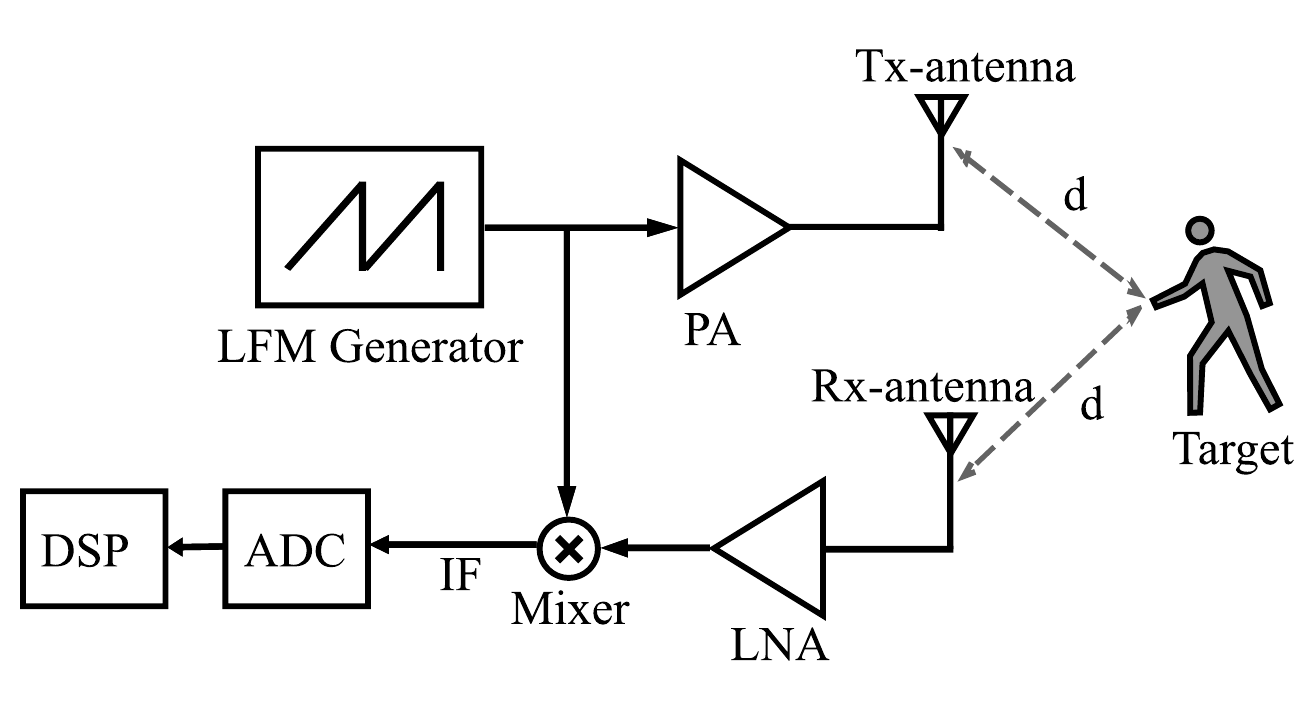}%
\label{BD}}
\hfil
\subfloat[]{\includegraphics[width=0.3\columnwidth,valign=b]{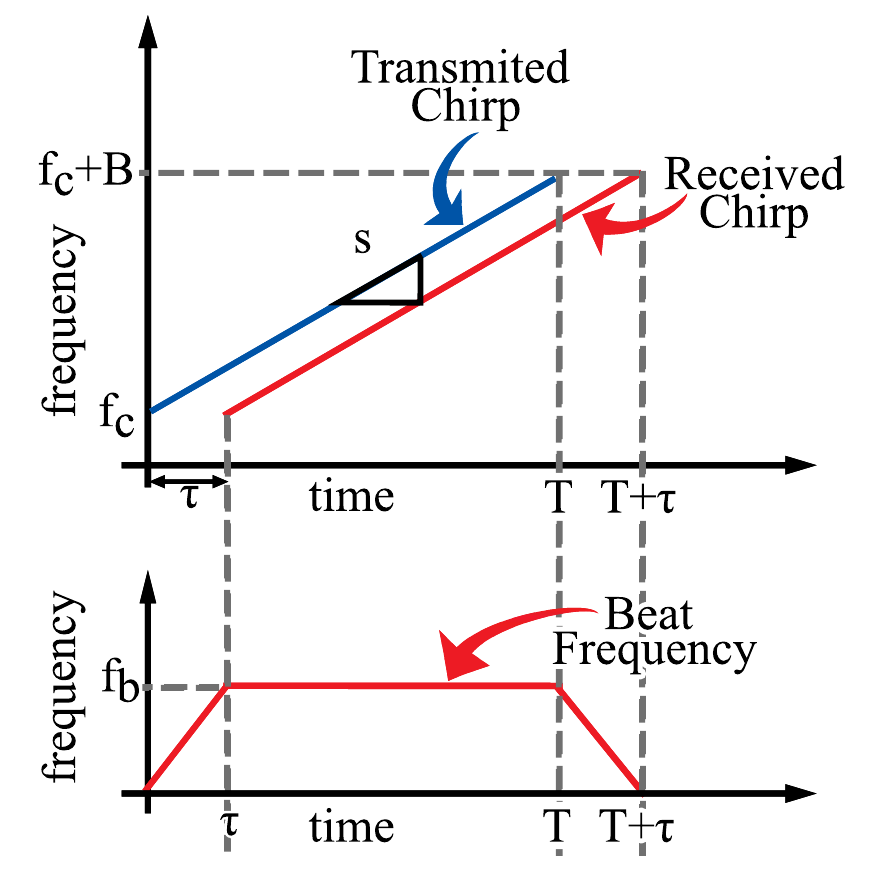}%
\label{DeChirp}}
\hfil
\subfloat[]{\includegraphics[width=0.3\columnwidth,valign=b]{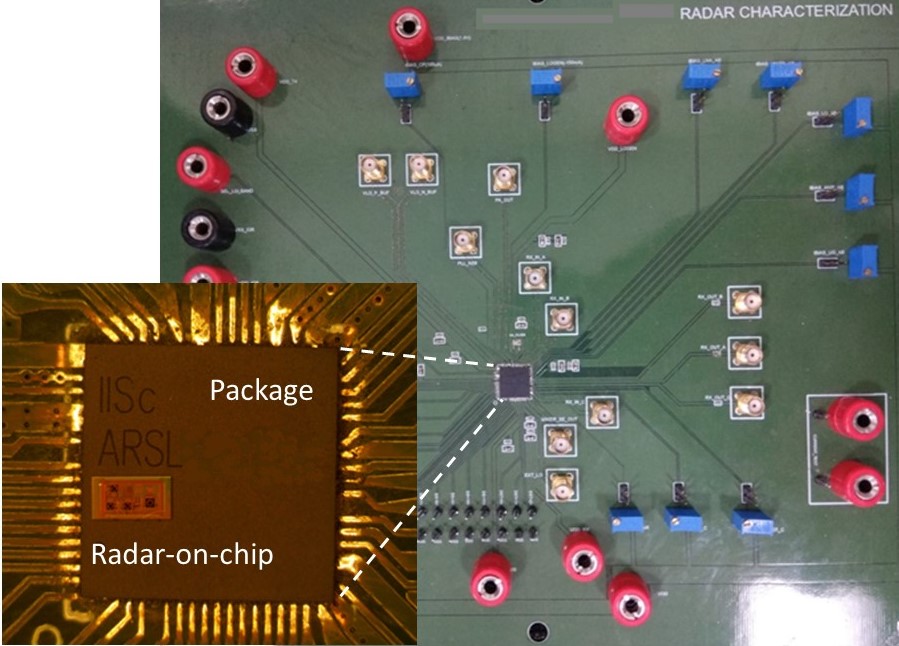}%
\label{chip}}
\caption{(a) Block diagram of an FMCW radar, (b) graphical representation of de-chirping, and (c) radar test board with a snapshot of the package and radar-on-chip (inset).}
\vspace{-0.4cm}
\label{radar}
\end{figure}

Fig.~\ref{radar}(a) shows the block diagram of an FMCW Radar system. For an FMCW radar with start frequency $f_c$, chirp bandwidth $B$ and modulation time $T$ with a target located at a distance $d$ from the radar, the transmitted LFM chirp can be expressed as: 
\begin{equation} \label{eq1} 
    S_{Tx} (t) = a_{0} \cos \left[2 \pi \left(f_{c} t + \frac{st^{2}}{2} \right) + \phi_{0} \right]  
\end{equation} 
where, $s$ is slope of the chirp defined as, s= $\frac{B}{T}$ and $\phi_0$ is the initial phase. After reflecting off a target situated at a distance $d$ from the Radar, the received signal is a time delayed replica of the transmitted chirp, given by: 
\begin{equation} \label{eq2}  
    S_{Rx} (t) = a_{1} \cos \left [2 \pi \left (f_{c} \left(t- \tau \right) + \frac{s\left(t- \tau \right)^{2}}{2} \right) +\phi_0 \right]   
\end{equation} 
The change in the amplitude incorporates the path loss and the radar cross section (RCS) of the target. The time delay ($\tau$) is the time taken by the EM wave to travel the round-trip distance, $2d$ given as $\tau=\frac{2d}{c} $.

The received signal is then mixed with an exact replica of the transmitted signal which is applied to the LO port of the mixer. The resulting waveform at the output port of the mixer is called the intermediate frequency (IF) signal expressed as 
\begin{equation} \label{eq3}  
    S_{IF} (t) = b \cos \left[2\pi \left (s\tau t + f_c\tau -\frac{s\tau^{2}}{2} \right)\right]
\end{equation} 
This process is termed de-chirping as illustrated in Fig.~\ref{radar}(b). For a single stationary target, the IF signal is a single tone with its frequency linearly proportional to the delay $\tau$ of the echo, and thus related to the beat frequency $f_b$ as 
\begin{equation} \label{eq4} 
    f_b = s\tau = \frac{2sd}{c}   
\end{equation} 

The target range can thus be determined from the beat frequency using (\ref{eq4}). The maximum detected range depends on the transmit power and the sampling frequency of the analog-to-digital converter (ADC), used to digitize the de-chirped signal. The range resolution of a radar is its ability to resolve two closely spaced objects. For two objects separated by a distance of $\Delta d$, the corresponding difference in their beat frequencies using (\ref{eq4}) can be written as
\begin{equation} \label{eq5}
    \Delta f_b = \frac{2s \Delta d}{c}   
\end{equation} 
Also, an observation window of T can resolve two frequencies that are separated by more than more than $1/T$ Hz, i.e., $\Delta f_b \geq 1/T$. 
Using (\ref{eq5}), we get 
\begin{equation} \label{eq6} 
    \Delta d \geq \frac{c}{2B}  
\end{equation} 
Thus, the range resolution of the FMCW radar depends directly on the bandwidth of the radar. 
\subsection{Radar On Chip} \label{roc}
In this paper, our experiments are based on a through wall radar on-chip (TWRoC)
developed in our institute \cite{news}.  The TWRoC (Fig. \ref{radar}(c)) was developed using a 180-nm CMOS technology integrating one transmitter and three receivers along with a fractional-N 
frequency synthesizer, which can generate complex radar signals including FMCW chirps.
In the FMCW mode, the radar operates from 2.05-2.6 GHz resulting in a 550 MHz bandwidth 
equivalent to a range resolution of 27 cm. Fig. \ref{radarroom}. is the pictorial representation of a compact TWRoC in action to determine the positions of targets of interest inside a room in presence of multiple stationary clutters.
\section{Antenna Design} \label{Antenna Design}
A radar-on-chip enables extremely portable, drone or robot mounted systems to explore various geophysical phenomena. Large form-factor antennas cannot be used in such systems. FMCW, which is widely used in automotive radars, is  a strong candidate for TWR and GPR. However, varying propagation for different frequencies and different building materials, impacts wideband FMCW operation. The design of the two antennas presented in this section was influenced by the need to support wideband FMCW radars. Full-wave EM simulations for the design were performed on CST Microwave Studio Suite 2021 \cite{cst}.
\subsection{Vivaldi Antenna} \label{Vivaldi Antenna}
\begin{figure}[b]
    \centering
    \includegraphics[width=\columnwidth,height=6.5cm,keepaspectratio]{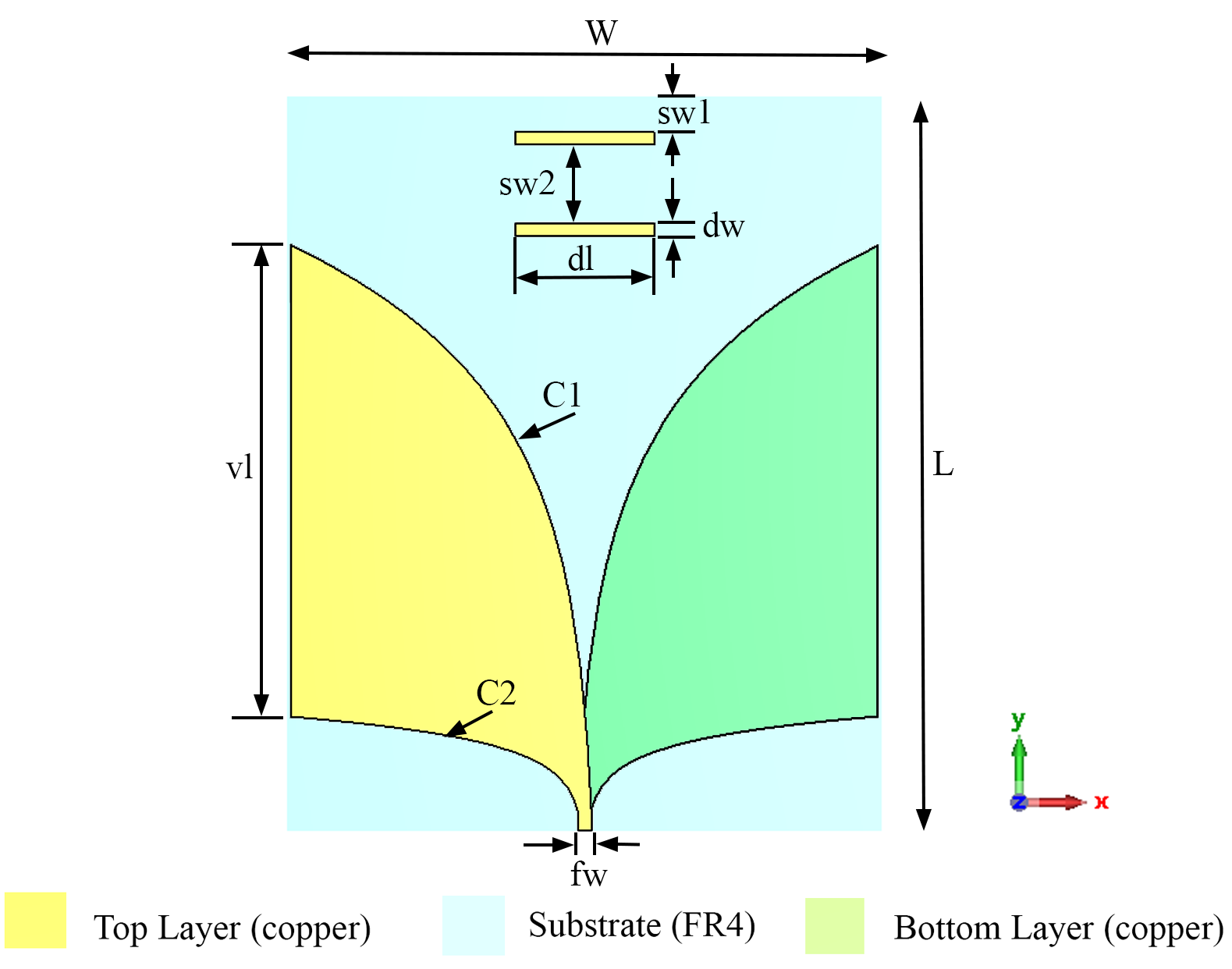}
    \caption{Schematic diagram of Vivaldi Antenna. Design parameters: W=150~mm, L=185~mm, v1=118.71~mm, sw1=9~mm, sw2=20~mm, fw=3.3~mm, dl=35~mm, and dw=3~mm.}
    \label{vdes}%
    \vspace{-0.4cm}
\end{figure}
The design of the Vivaldi antenna with 2 directors for enhanced gain is shown in Fig.~\ref{vdes}. The antenna is designed on an FR-4 substrate of thickness 1.6~mm with relative permittivity 4.3 and loss tangent 0.025. 
The inner edge and outer edge of both the arms of the antenna are exponential and each arm is metallized on either side of the substrate, flaring in the opposite direction to form a tapered slot. EM waves travel along the inner edges of the flared aperture and couple with each other to produce radiation. According to conventional theory \cite{gazit1988improved}, the lower cutoff frequency of a Vivaldi antenna is determined by the width of antenna aperture, which can be expressed as $\lambda_{cutoff} = 2W$. The exponential curve for the inner edge C1 is defined as;
	\begin{align}\label{eq:veq1}
        X(t)= \text{fw} - 0.5 \text{fw} e^{(a1(t-5)}+\frac{W1}{2}	, ~~
        Y(t)= t
	\end{align}
where $t \in$ [5, 148], $fw$ is the width of feeding microstrip, and  $a1$ = 0.027 is the rate of exponential curve for C1, and for the outer edge C2 is defined as;
	\begin{align}\label{eq:veq2}
        X(t)= - 0.5 \text{fw} e^{a2(t-5)}+\frac{W1}{2}, ~~
        Y(t)= t
    \end{align}
where t $\in$ [5, 29] and $a2$ = 0.16 is the rate of exponential curve for C2.

\begin{figure}
\centering
\subfloat[]{\includegraphics[width=0.45\columnwidth,valign=b]{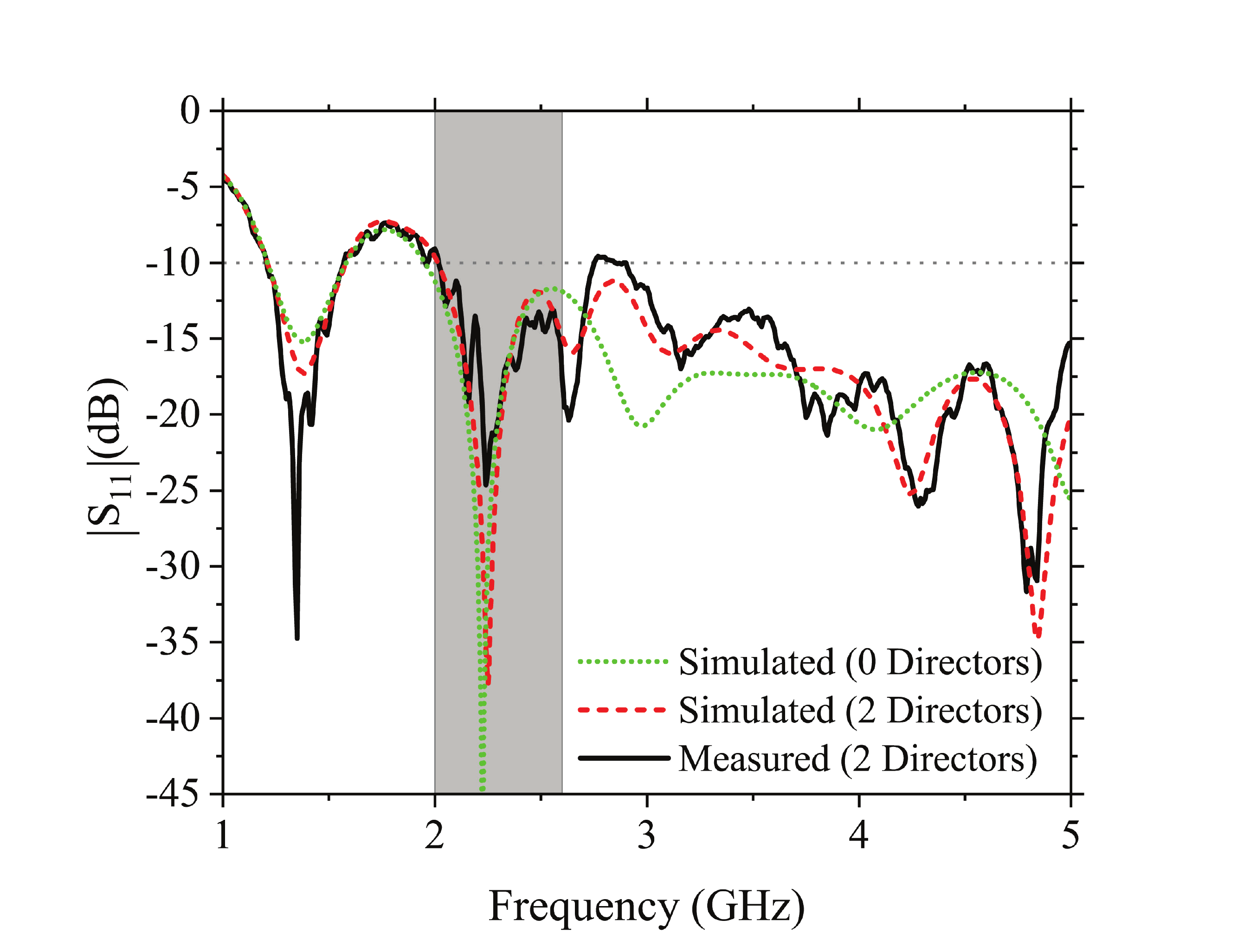}%
\label{vspara}}
\hfil
\subfloat[]{\includegraphics[width=0.45\columnwidth,valign=b]{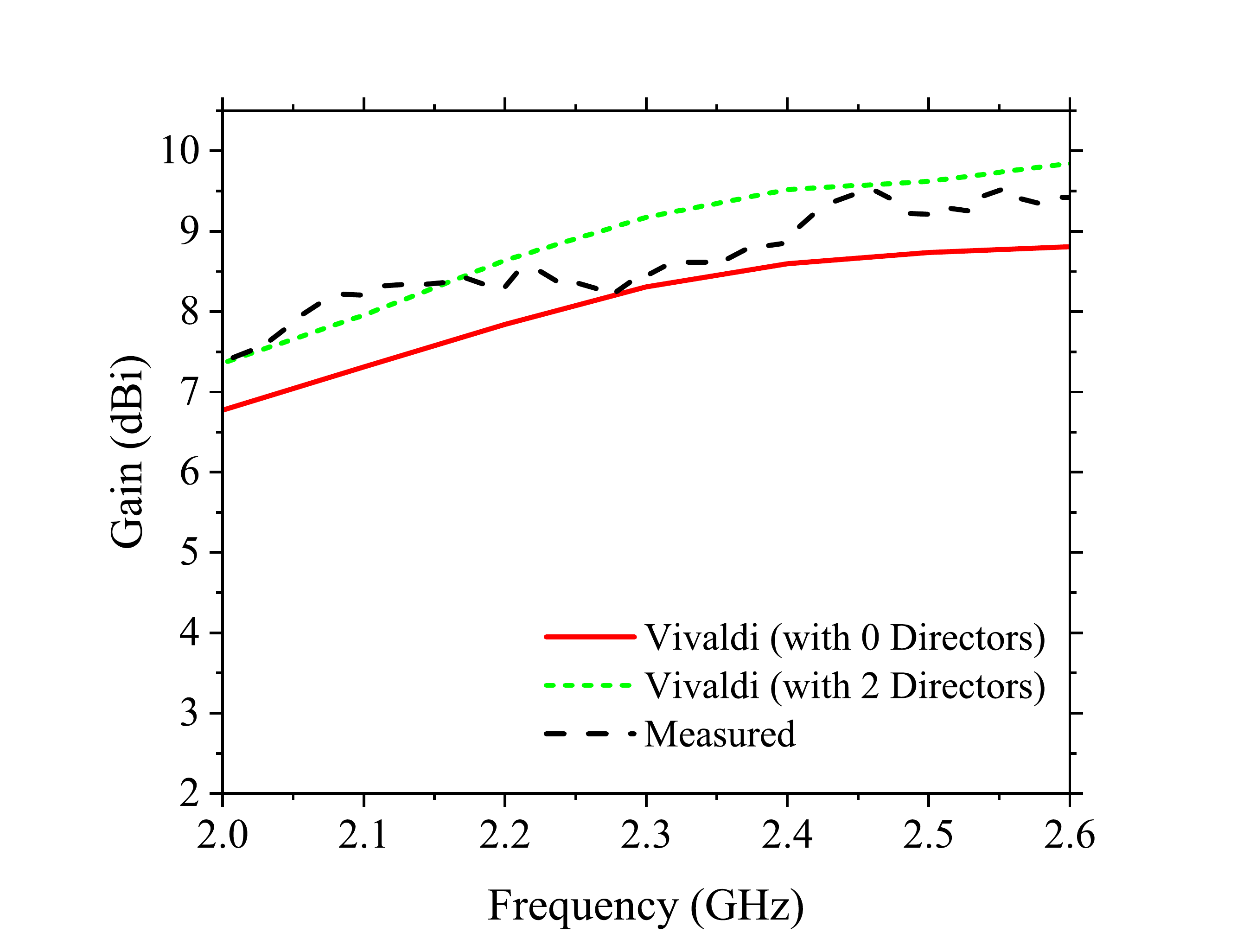}%
\label{vgain}}
\caption{(a) Frequency variation of $|S_{11}|$ for the Vivaldi antenna design in Fig.~\ref{vdes}, and (b) frequency variation of peak gain for the Vivaldi antenna for different number of directors.}%
\vspace{-0.5cm}
\label{vsg}
\end{figure}
\begin{figure}
\centering
\subfloat[]{\includegraphics[width=0.33\columnwidth,valign=t]{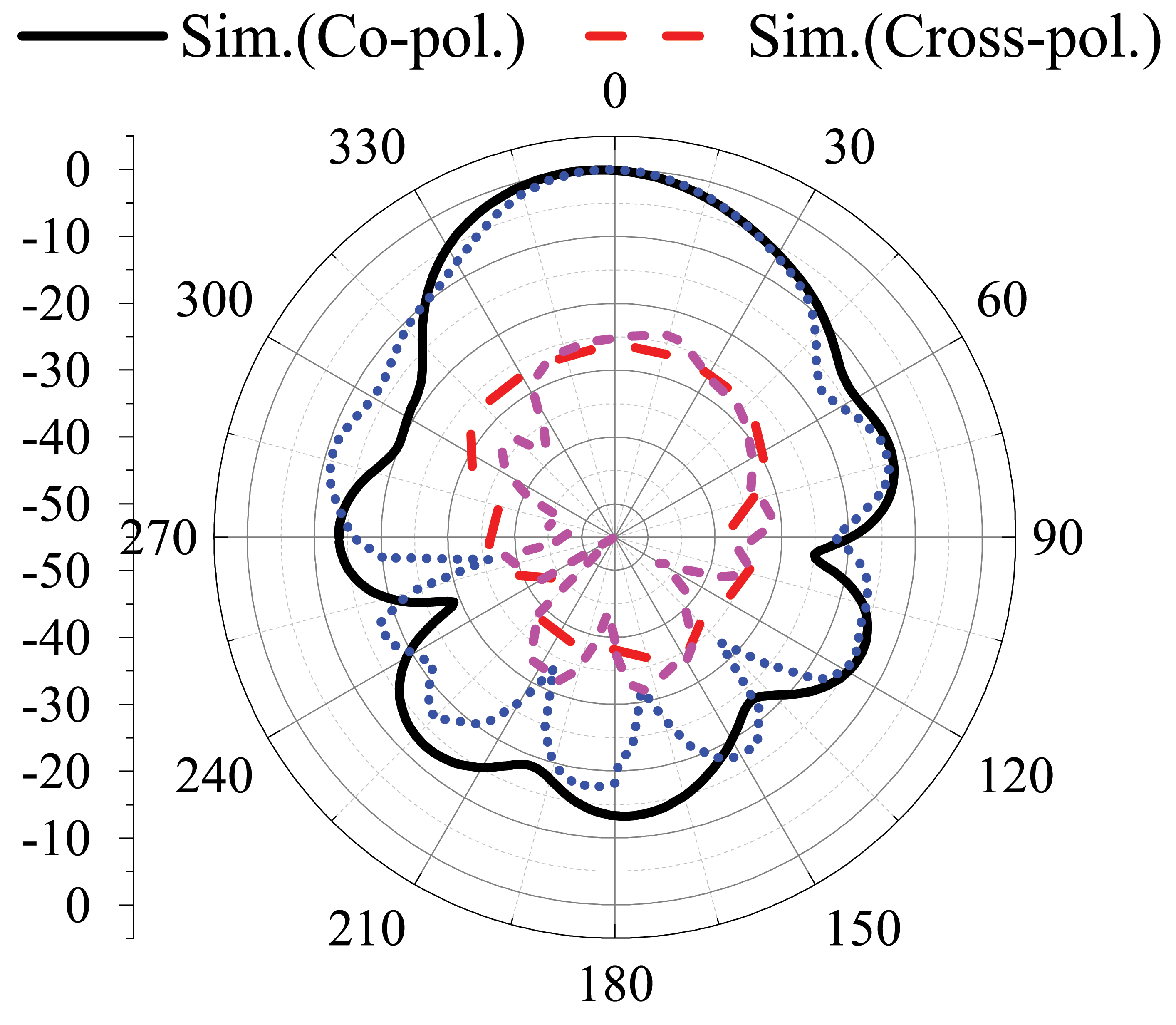}%
\label{Vivrad1}}
\hfil
\subfloat[]{\includegraphics[width=0.35\columnwidth,valign=t]{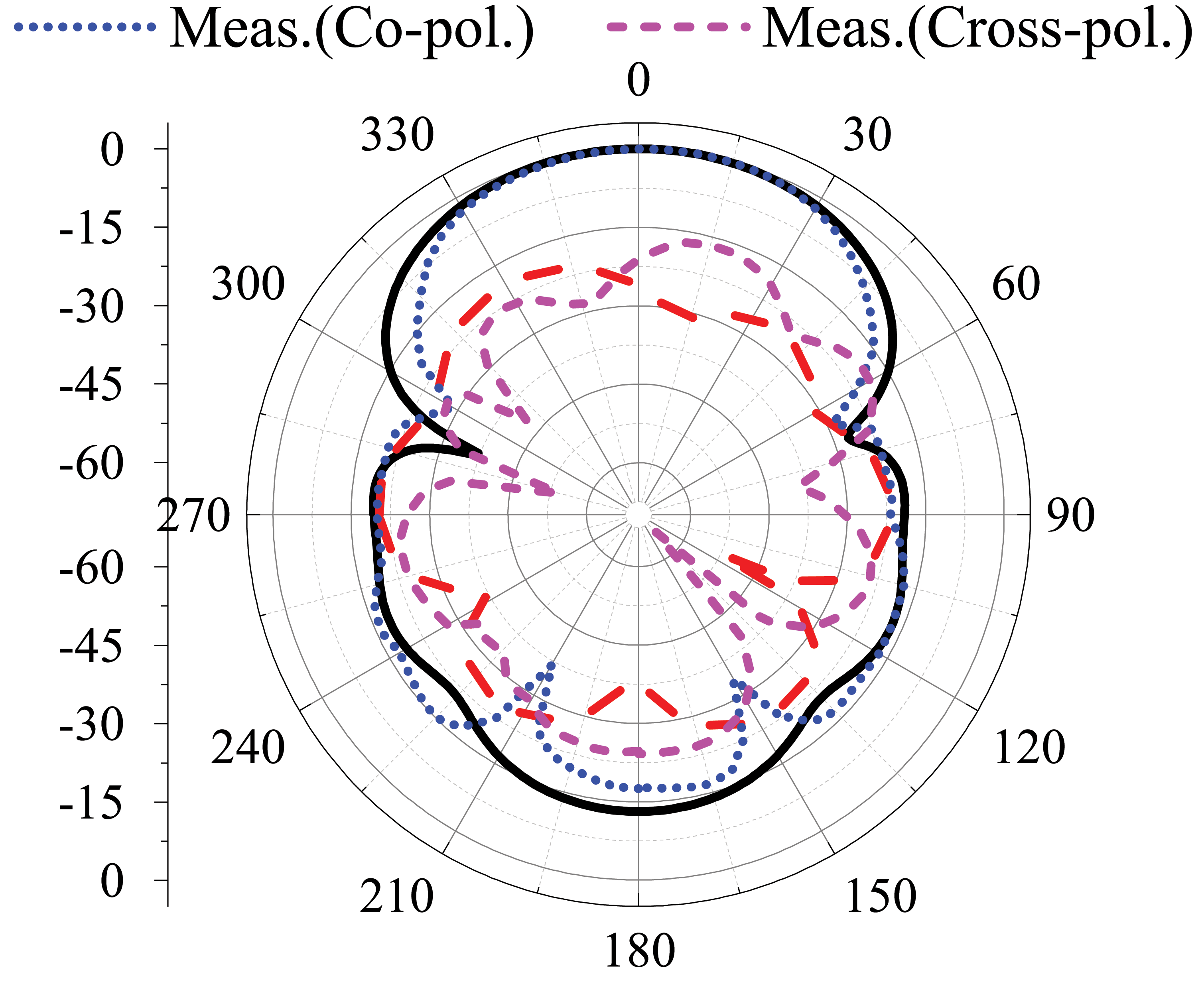}%
\label{Vivrad2}}
\caption{Radiation pattern (a) E-plane and (b) H-plane for the proposed Vivaldi antenna at 2.4~GHz.}%
\vspace{-0.5cm}
\label{Vivrad}
\end{figure}
\begin{figure}\centering
    \includegraphics[width=\columnwidth,height=4cm,keepaspectratio]{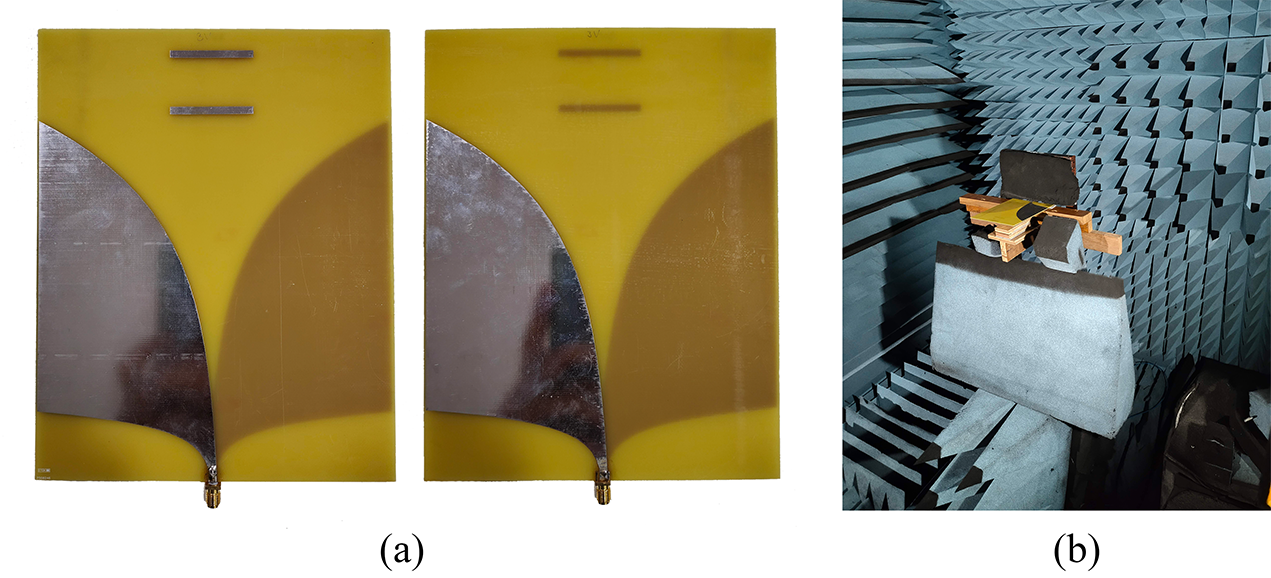}
    \caption{Fabricated prototype of the Vivaldi antenna and test setup in the anechoic chamber.}
    \vspace{-0.6cm}
    \label{Vivmes}
\end{figure}
The antenna is designed to be fed through a 50 $\Omega$ SMA connector followed by a gradual transition consisting of an exponentially contoured ground plane and a microstrip line, shown in Fig.~\ref{vdes}. Full 3D wave EM simulations for the design were performed on CST Microwave Studio Suite 2021 \cite{cst}. Fig.~\ref{vsg}(a) shows the $S_{11}$ plots for the designed Vivaldi antenna. As shown in Fig.~\ref{vsg}(b), with 0 directors, the gain of the antenna varies from 6.8 to 8.8~dBi. To enhance the gain of the Vivaldi antenna, 2 directors are added between the inner edges. Consequently, the gain increases by 1~dB.

The radiation pattern plots for the E- and H-planes are shown in Fig.~\ref{Vivrad}. The antenna has a directivity of 10.44~dBi and front-to-back ratio of 14.44~dB at 2.4~GHz. The fabricated antenna and the measurement setup is shown in Fig.~\ref{Vivmes}. The measured result is obtained using a Keysight N5230A PNA series network analyzer, and the influence of the coaxial line is eliminated by calibration. It is observed that simulation results match the experimental results reasonably well.

\subsection{Proposed Quasi-Yagi Antenna} \label{Quasi-Yagi Antenna}
\begin{figure}\centering
    \includegraphics[width=\columnwidth,height=8cm,keepaspectratio]{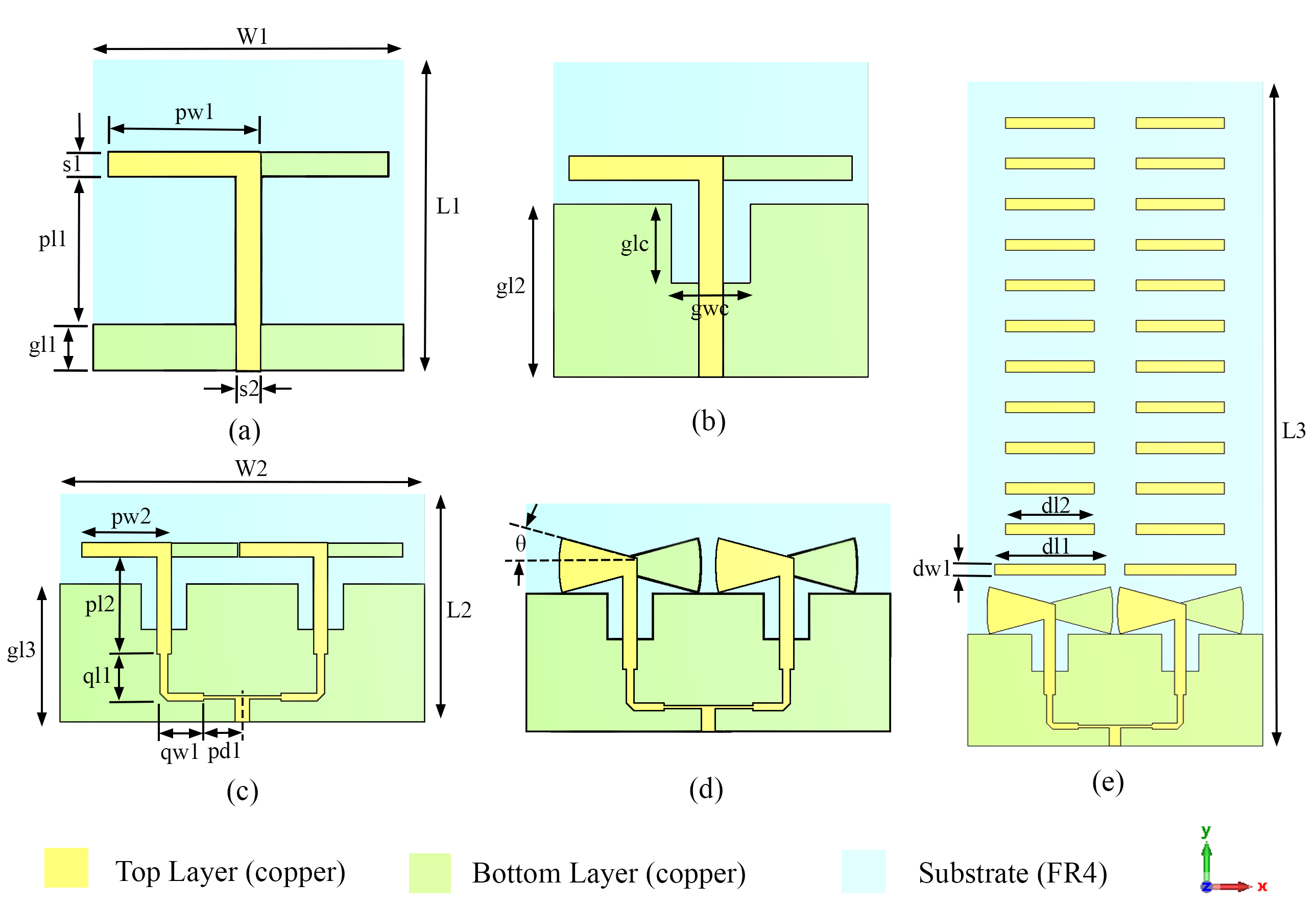}
    \caption{Design process for proposed quasi-Yagi antenna (a) Stage-I: initial design, (b) Stage-II: initial design with rectangular slot introduced in the ground, (c) Stage-III: two element array for the design in Stage-II, (d) Stage-IV: design in Stage-III with flared dipoles and (e) Stage-V: final design of the two element array for quasi-Yagi antenna with modified ground, flared dipoles and 12 directors .}%
    \vspace{-0.4cm}
    \label{QYdes}
\end{figure}

This section describes the design procedure for the proposed compact quasi-Yagi antenna for the TWR radar-on-chip (section~\ref{roc}). Fig.~\ref{QYdes} illustrates the evolution of the designed antenna structure at various stages.
A microstrip-fed printed directive dipole antenna operating at 2.34~GHz with an impedance bandwidth (IBW) of 13.24\% ($S_{11}<$~-10~dB) is chosen as the initial design (Fig.~\ref{QYdes}(a)). The directive dipole consists of an x-directed printed dipole antenna with a partial ground plane. The printed dipole is excited by broadside-coupled co-planar strip-lines (BC-CPS). The partial ground plane acts as the reflector, which is essential for the directive operation of the antenna. The antenna is designed on a 1.6~mm thick FR-4 epoxy substrate with relative permittivity ($\epsilon_{r}$) of 4.3 and loss tangent (tan~$\delta$) of 0.025. 
\begin{table}
\centering
\caption{\textsc{List of Design Parameters for quasi-Yagi Antenna}}
\label{tab:QYp}
\resizebox{0.8\columnwidth}{!}{%
\begin{tabular}{@{}llllllll@{}}
\toprule
\multicolumn{1}{c}{\textbf{Parameter}} & \multicolumn{1}{c}{\textbf{\begin{tabular}[c]{@{}c@{}}Value \\ (mm)\end{tabular}}} &
\multicolumn{1}{c}{\textbf{Parameter}} & \multicolumn{1}{c}{\textbf{\begin{tabular}[c]{@{}c@{}}Value \\ (mm)\end{tabular}}} &
\multicolumn{1}{c}{\textbf{Parameter}} & \multicolumn{1}{c}{\textbf{\begin{tabular}[c]{@{}c@{}}Value \\ (mm)\end{tabular}}} & \multicolumn{1}{c}{\textbf{Parameter}} & \multicolumn{1}{c}{\textbf{\begin{tabular}[c]{@{}c@{}}Value \\ (mm)\end{tabular}}} \\ \midrule
W1     & 40         & L1         & 40        & pw1           & 19.55     & pl1        & 19     \\                                  s1     & 3.1        & s2         & 3.1       & gl1           & 6         & gl2        & 22     \\                                  glc    & 10         & gwc        & 10        & W2            & 80        & L2         & 50     \\
pw2    & 19.55      & pl2        & 21.25     & gl3           & 30.25     & ql1        & 10.33  \\                                  qw1    & 9.5        & pd1        & 8.5       & dw1           & 3         & dl1        & 30     \\                                  dl2    & 24         & $\theta$   & $15^{\circ}$  &           &           &            &        \\ \bottomrule
\end{tabular}%
}\vspace{-0.1cm}
\end{table}
\begin{figure}
\centering
\subfloat[]{\includegraphics[width=0.45\columnwidth]{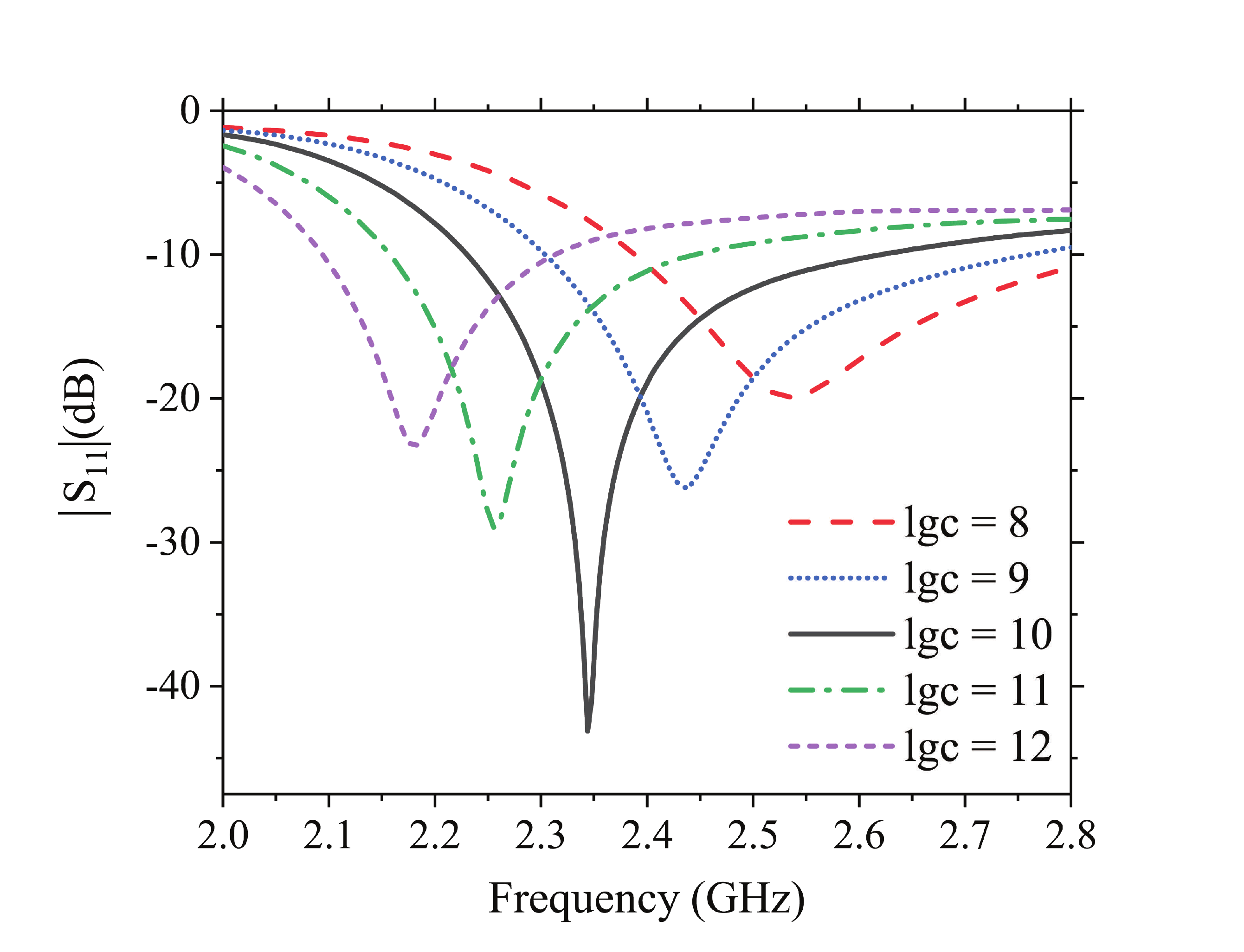}%
\label{lgc}}
\hfil
\subfloat[]{\includegraphics[width=0.45\columnwidth]{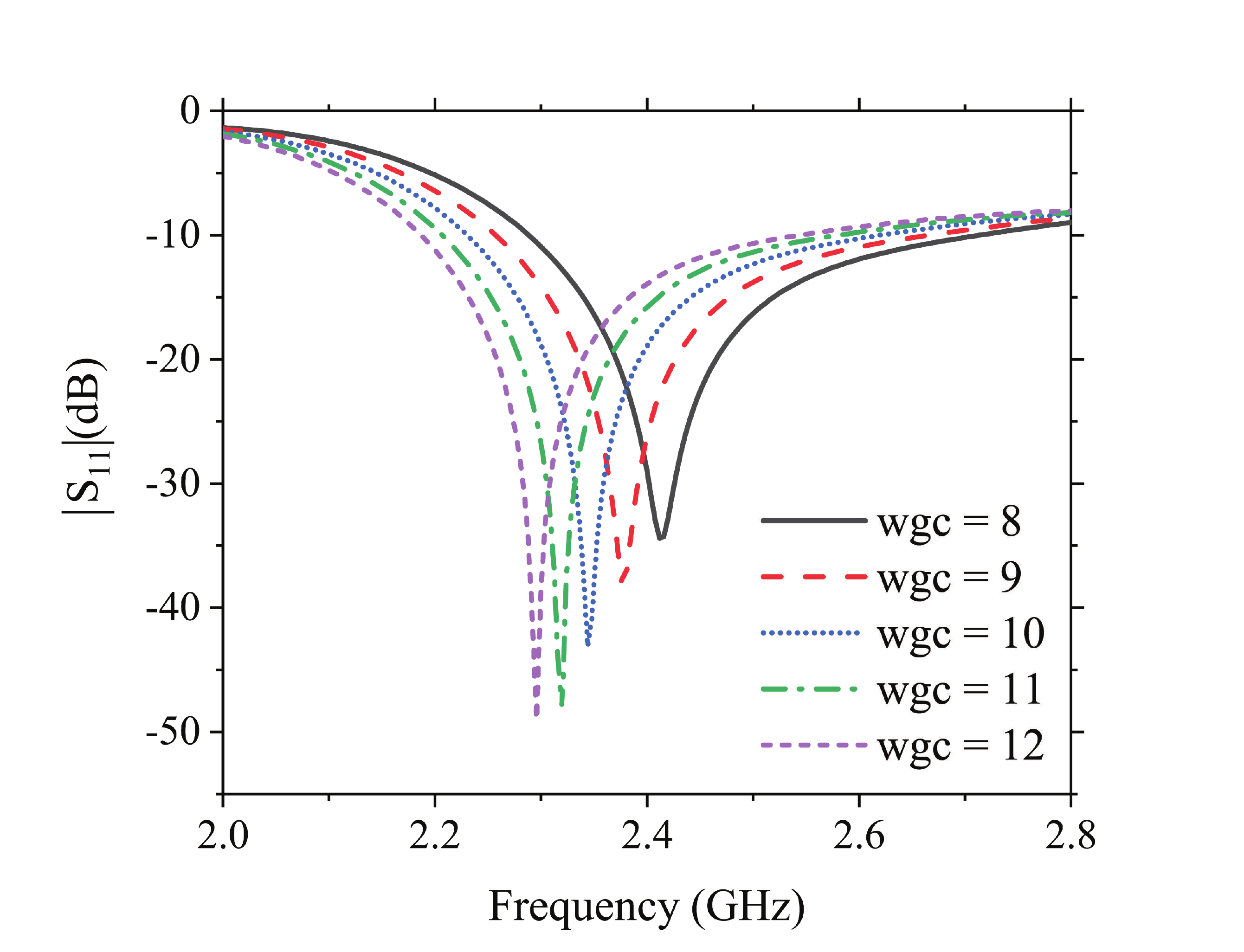}%
\label{wgc}}
\caption{Frequency variation of $|S_{11}|$ with slot parameters: (a) lgc (mm) and, (b) wgc (mm).}
\label{slot}
\vspace{-0.4cm}
\end{figure}

For a length of 1.25$\lambda$ (46~mm) the dipole radiates at 2.34~GHz. To reduce the size of the dipole a rectangular slot is introduced near the BC-CPS (Fig.~\ref{QYdes}(b)). This modification of the ground plane shifts the centre frequency. The introduced slot has two parameters (glc and gwc) which can be tuned to obtain matching at the desired frequency. A parametric study to demonstrate the impact of glc and gwc variation, is shown in Fig.~\ref{slot}.
\begin{figure}
\centering
\subfloat[]{\includegraphics[width=0.45\columnwidth]{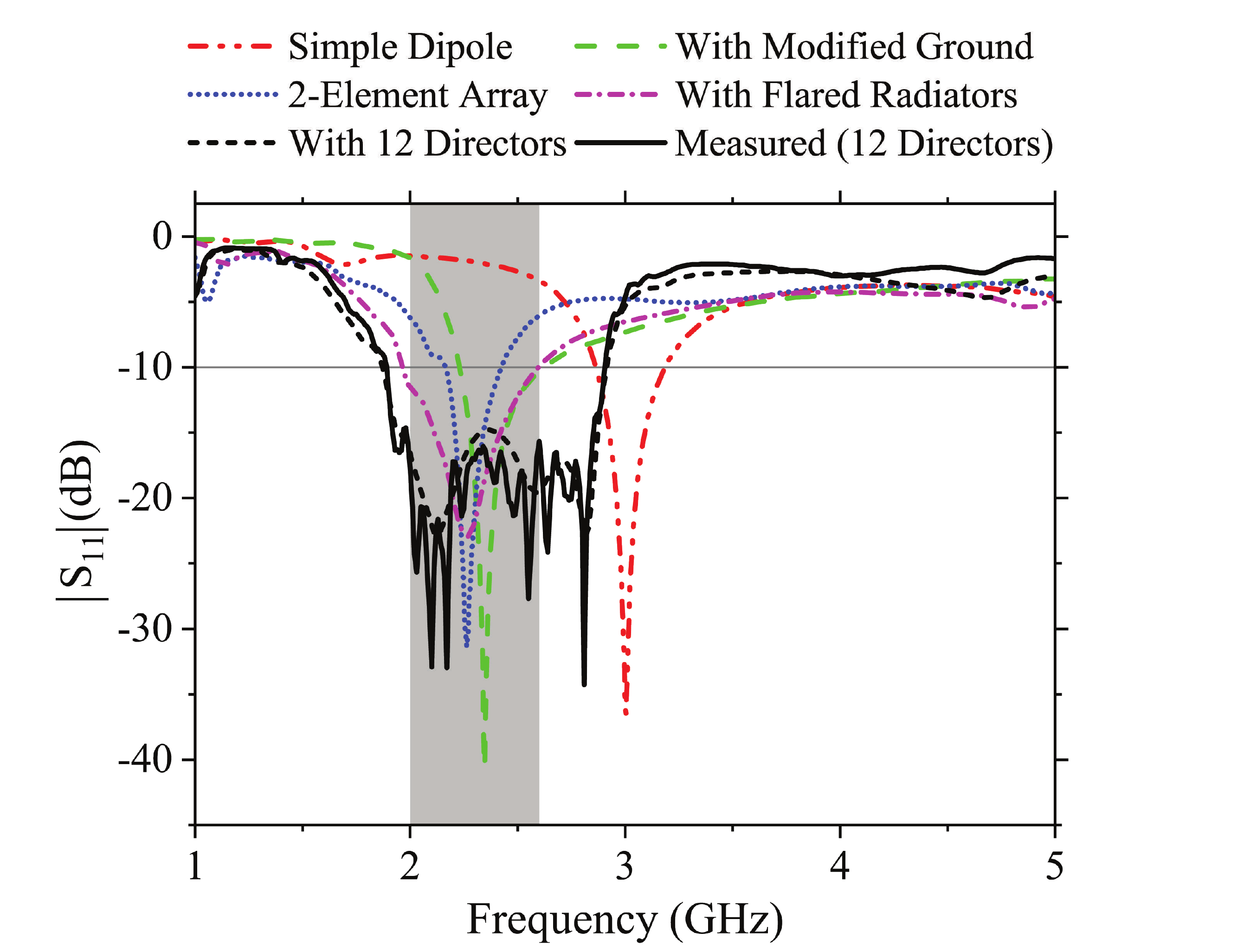}%
\label{QYspara}}
\hfil
\subfloat[]{\includegraphics[width=0.45\columnwidth]{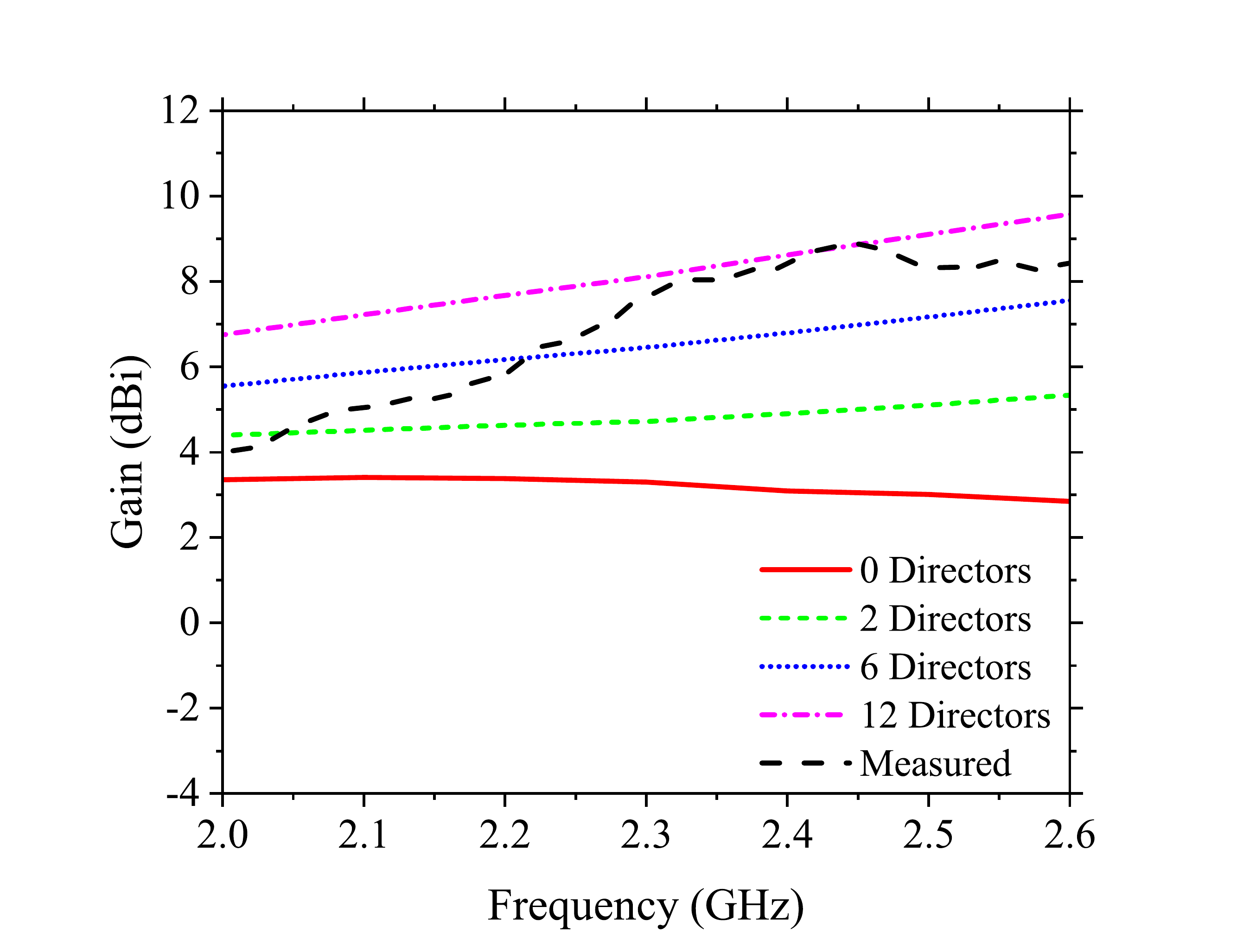}%
\label{QYgain}}
\caption{(a) Frequency variation of $|S_{11}|$ for the proposed quasi-Yagi antenna design in Fig.~\ref{QYdes}, and (b) frequency variation of peak gain for the proposed quasi-Yagi antenna for different number of directors.}%
\vspace{-0.4cm}
\label{QYsg}
\end{figure}
The dimensions of the slot were optimised (glc = gwc = 10~mm) to achieve a smaller dipole length (=$0.97\lambda$) for the same resonant frequency. The design was then used to obtain a 2-element array (Fig.~\ref{QYdes}(c)) and to increase the bandwidth, the dipoles were flared at an angle of $\theta = 15^{\circ}$ (Fig.~\ref{QYdes}(d)). Flaring in printed microstrip dipole antenna has been shown to improve the impedance bandwidth without adversely affecting the efficiency \cite{dey1991bandwidth}.  Upon flaring the dipoles, the IBW increased from 13.24\% to 27.35\%. 
\begin{figure}[!t]
\centering
\subfloat[]{\includegraphics[width=0.35\columnwidth,valign=c]{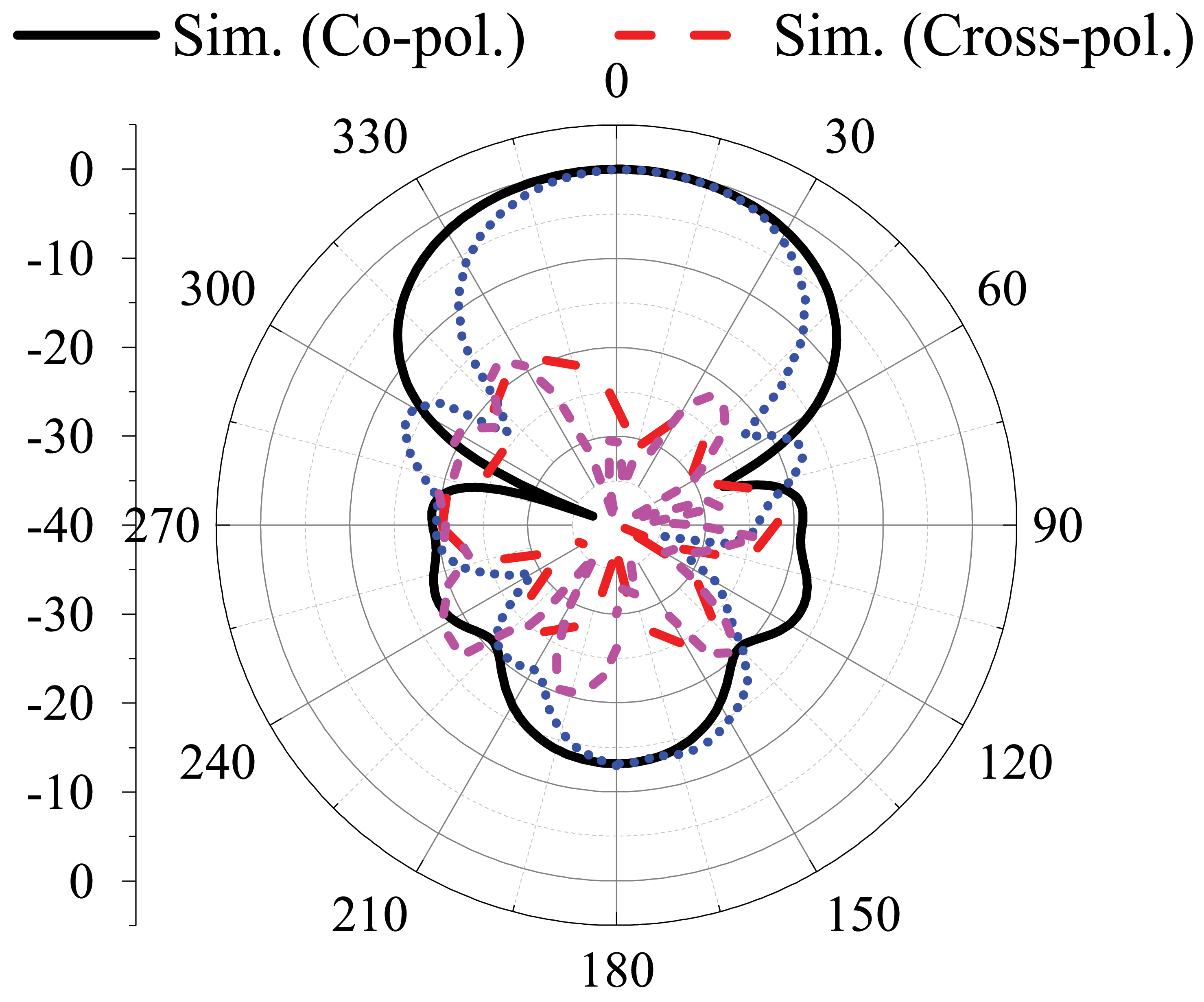}%
\label{QYrad1}}
\hfil
\subfloat[]{\includegraphics[width=0.35\columnwidth,valign=c]{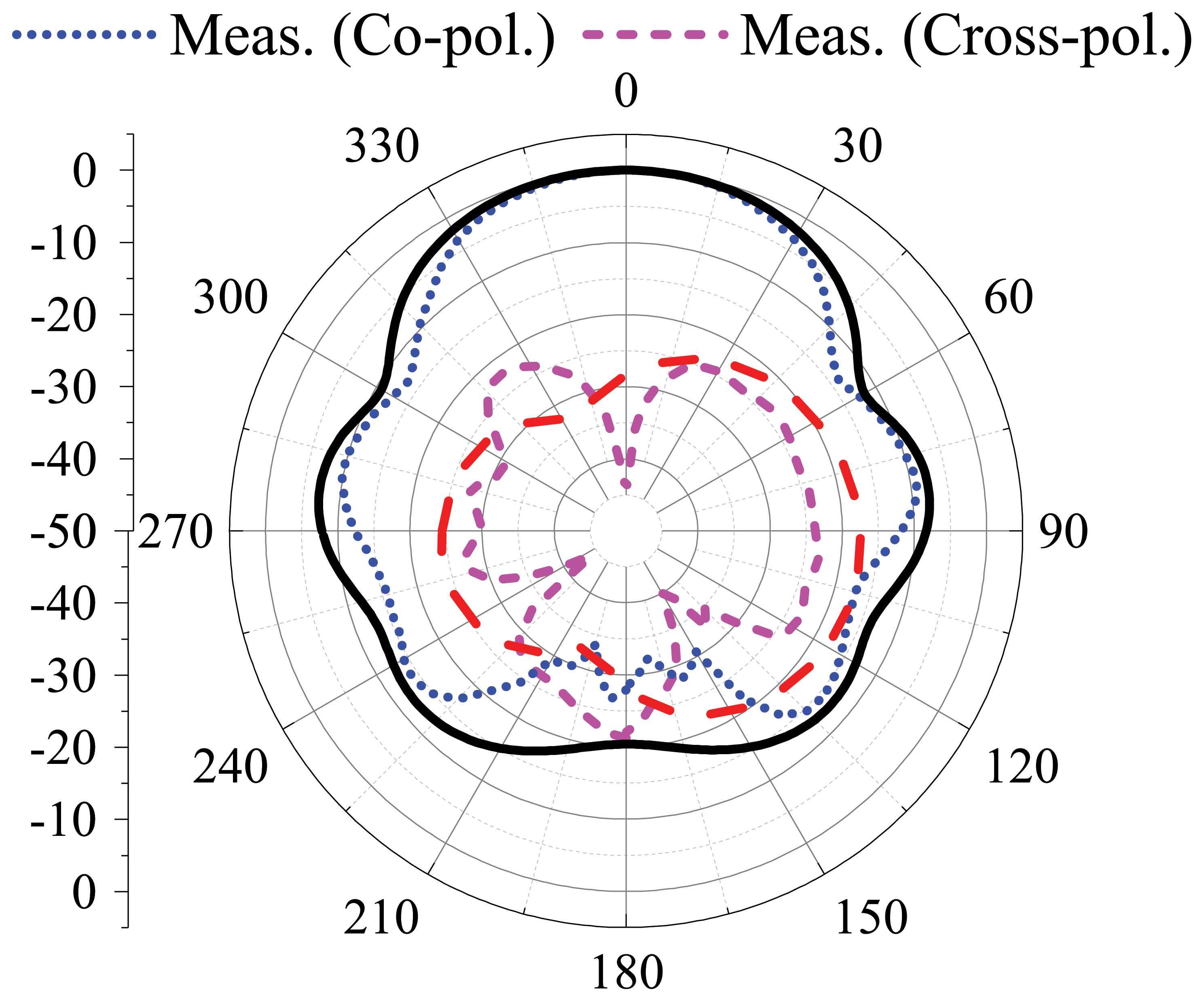}%
\label{QYrad2}}
\caption{Radiation pattern (a) E-plane and (b) H-plane for the proposed quasi-Yagi antenna at 2.4~GHz.}
\label{QYrad}%
\vspace{-0.4cm}
\end{figure}
\begin{figure}\centering
    \includegraphics[width=\columnwidth,height=5cm,keepaspectratio]{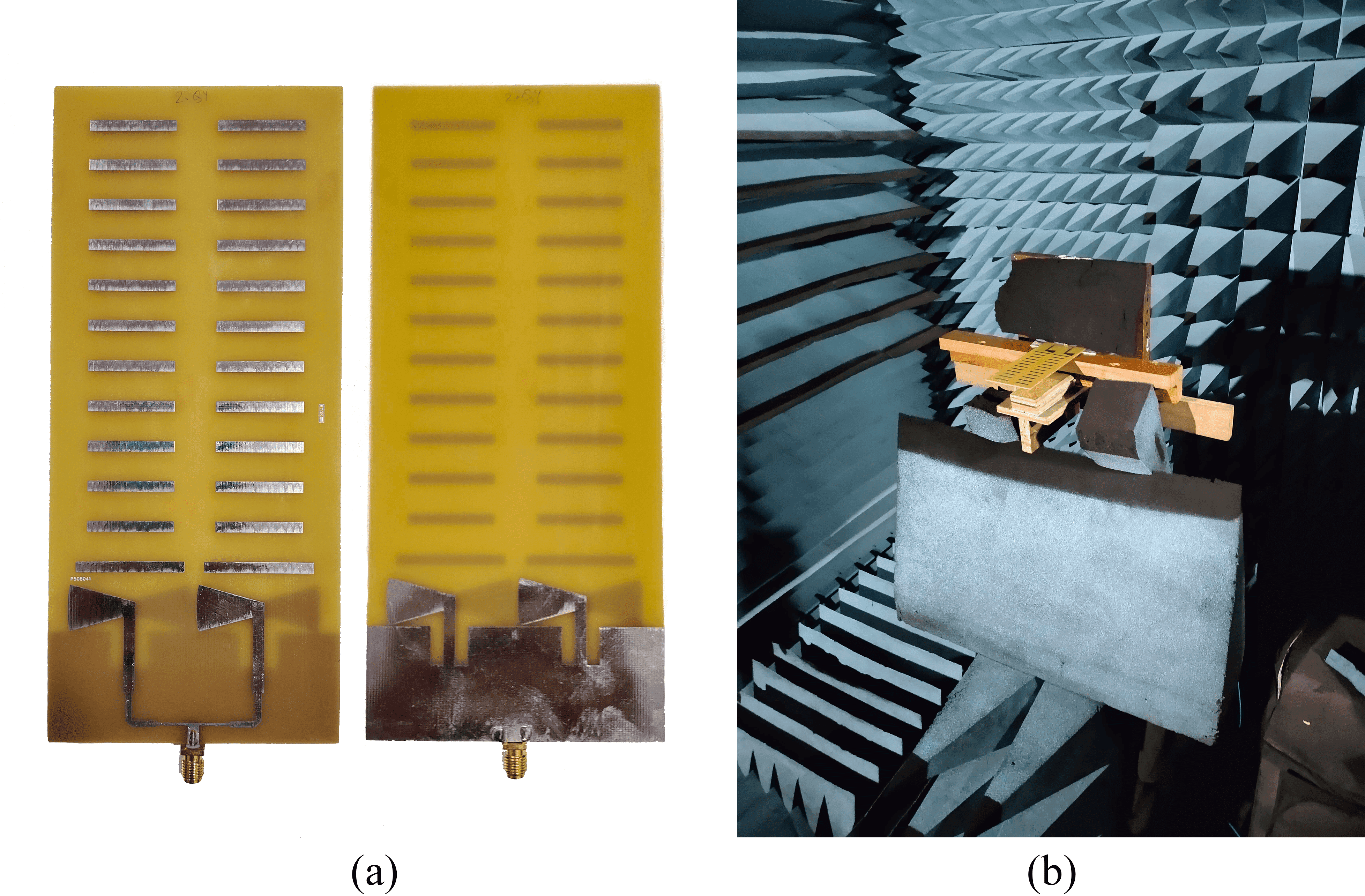}
    \caption{Fabricated prototype of proposed quasi-Yagi antenna and test setup in the anechoic chamber.}
    \label{QYmes}%
    \vspace{-0.2cm}
\end{figure}
Next, to improve the gain of the antenna, directors were added (Fig.~\ref{QYdes}(d)). Starting with no directors, the number of directors was increased gradually. The gain of the antenna at 2.4~GHz for 0, 2, 6 and 12 directors was observed to be 3.09~dBi, 4.90~dBi, 6.80~dBi and 8.62~dBi, respectively. Due to the trade-off between the gain and the size of the antenna, the number of directors was chosen to be 12. The final design offers a high IBW of 44.62\%. The $|S_{11}|$ versus frequency plots for the antenna at each design stage are shown in Fig.~\ref{QYsg}(a) and the gain plot for increasing number of directors is shown in Fig.~\ref{QYsg}(b). The radiation patterns for the E-and H-planes at 2.4~GHz, are shown in Fig.~\ref{QYrad}. The antenna has a measured gain of 8.7~dBi and front-to-back ratio of 25.76~dB at 2.4~GHz.
\begin{table}
\centering
\caption{\textsc{Comparison with previously reported compact planar quasi-Yagi antennas}}
\label{tab:CompQY}
\resizebox{\textwidth}{!}{%
\begin{tabular}{@{}ccccccccc@{}}
\toprule
\textbf{Design}                                                           & \textbf{\begin{tabular}[c]{@{}c@{}}Bandwidth\\ (GHz)\end{tabular}} & \textbf{\begin{tabular}[c]{@{}c@{}}Frequency\\ (GHz)\end{tabular}} & \textbf{\begin{tabular}[c]{@{}c@{}}Gain\\ (dBi)\end{tabular}} & \textbf{\begin{tabular}[c]{@{}c@{}}Relative  per\\ -mittivity ($\epsilon_{r}$)\end{tabular}} & \textbf{\begin{tabular}[c]{@{}c@{}}Size \\ ($\lambda_{0} \times \lambda_{0}$)\end{tabular}} & \textbf{\begin{tabular}[c]{@{}c@{}}Size \\ ($\lambda_{g} \times \lambda_{g}$)\end{tabular}} & \textbf{Description}    & \textbf{\begin{tabular}[c]{@{}c@{}}Figure of\\  Merit\end{tabular}}                                                                                       \\ \midrule
\cite{Silva2017Quasi}                                               & 2.3-3.03                                                          & 2.45                                                               & -                                                       & 4.65                                                                                                         & 0.7 x 0.97                                                                                & 1.4 x 1.94                                                                                 & Microstrip to Co-planar Stripline                                                                                & - \\
\cite{Aeini2017CompactWQ}                                               & 1.43-3.97                                                          & 2.7                                                                & 3.4-5.2                                                       & 4.4                                                                                                         & 0.675 x 0.67                                                                                & 2.37 x 2.4                                                                                  & Spiral metamaterial resonators                                                                                 & 2.65 \\
\cite{Rezaeieh2017MiniaturizedPY}                                       & 0.69-1.12                                                          & 1                                                                  & 3-5.5                                                         & 4.4                                                                                                         & 0.4 x 0.33                                                                                  & 1.58 x 1.22                                                                                 & Folded reflector                                                                                             & 1.08  \\
\cite{4718026}                                       & 2.36-2.55                                                          & 2.4                                                                  & 5.1                                                         & 10.2                                                                                                         & 0.46 x 0.69                                                                                  & 1.21 x 1.8                                                                                 & Backed by a metal reflector                                                                                              & 0.26 \\
\cite{Wu2014BandwidthEO}                                                & 3.6-10.25                                                          & 6.93                                                               & 4.5-8.3                                                       & 4.4                                                                                                         & 0.69 x 0.785                                                                                & 2.52 x 2.87                                                                                 & Extended stubs in ground plane                                                                              & 2.70    \\
\cite{4237129}                                                & 4.9-5.3                                                          & 5.2                                                               & 13                                                       & 2.2                                                                                                         & 2 x 2.38                                                                                & 4.18 x 5.23                                                                                 & \begin{tabular}[c]{@{}c@{}}Coupling energy to additional \\  patches \end{tabular}                                                                                 & 1.39 \\
\cite{7206518}                                                & 51-70                                                          & 60                                                               & 10.5-11.7                                                       & 3.38                                                                                                         & -                                                                                & 3.3 x 3.5                                                                                 & Ladder-like directors                                                                               & 3.55  \\
\cite{Liu2019ACP}                                                      & 1.98-2.69                                                          & 2.35                                                               & 5.7                                                           & 2.94                                                                                                        & 0.54 x 0.38                                                                                 & 1.67 x 1.16                                                                                 & \begin{tabular}[c]{@{}c@{}}DSPSL director, reflector, and \\ offset DSPSL parasitic element\end{tabular} & 1.11 \\
\cite{Yang2020CompactQA}                                                & 8.7-10.25                                                          & 9.55                                                               & 7.3  $\pm$ 0.75                                                  & 3.55                                                                                                        & 1.21 x 1.55                                                                                 & -                                                                                           & \begin{tabular}[c]{@{}c@{}}Cylinder dielectric resonator \\ and split-ring resonator\end{tabular}         & 0.87     \\
\cite{Hwang2019QuasiYagiAA}                                             & 26.3 - 29.75                                                       & 28                                                                & 5.51                                                          & 3.2                                                                                                         & 0.47 x 0.47                                                                                 & 1.49 x 1.49                                                                                 & Planar folded dipole topology                                                                           & 0.44       \\
\cite{Kim2018DesignAA}                                                  & 4.67 -   9.89                                                      & 8                                                                  & 6.46                                                          & 4.5                                                                                                         & 0.88 x 0.87                                                                                 & 3.29 x 3.27                                                                                 & Two parasitic directors                                                                                  &2.89      \\
\cite{Ashraf2020DesignAD}                                               & 2.6-4.6                                                            & 3.45                                                               & 8$\sim$9                                                      & 4.4                                                                                                         & 0.69 x 1.61                                                                                 & 2.46 x 5.75                                                                                 & \begin{tabular}[c]{@{}c@{}}Elliptically shaped \\ coupled-directive element\end{tabular}                 & 3.65      \\
\cite{5159418}                                               & 5.2-7.1                                                            & 5.5                                                               & 4.1                                                     & 3.38                                                                                                         & 0.52 x 0.52                                                                                 & 1 x 1                                                                                 & \begin{tabular}[c]{@{}c@{}}Microstrip-to-coplanar transition \\ using artificial transmission lines\end{tabular}              & 0.88         \\
\textbf{Fig.~\ref{QYdes} (b)}                                                       & \textbf{2.23-2.61}                                                 & \textbf{2.34}                                                      & \textbf{3-3.2}                                                & \textbf{4.3}                                                                                                & \textbf{0.31 x 0.31}                                                                        & \textbf{1.128 x 1.128}                                                                      & \textbf{Ground slot}                                                                                    & \textbf{0.34 }      \\
\textbf{\begin{tabular}[c]{@{}c@{}}Fig.~\ref{QYdes} (b) \\ (modified)\end{tabular}} & \textbf{2.16-2.6}                                                  & \textbf{2.34}                                                      & \textbf{4.16-5.65}                                            & \textbf{4.3}                                                                                                & \textbf{0.31 x 0.468}                                                                       & \textbf{1.128 x 1.689}                                                                      & \textbf{Ground slot, 2 directors}                                                                      & \textbf{0.68 }       \\
\textbf{Fig.~\ref{QYdes} (e)}                                                       & \textbf{1.87-2.91}                                                 & \textbf{2.34}                                                      & \textbf{6.8-9.8}                                              & \textbf{4.3}                                                                                                & \textbf{0.626 x 1.175}                                                                      & \textbf{2.22 x 4.17}                                                                        & \textbf{\begin{tabular}[c]{@{}c@{}}Ground slot, flared radiators \\ and 12 directors\end{tabular}}        & \textbf{4.24 }    \\ \bottomrule
\end{tabular}%
}
\end{table}
A comparison between the proposed and previously reported compact planar quasi-Yagi antennas is shown in Table~\ref{tab:CompQY}. The gain bandwidth product for the antennas is defined as the figure of merit. It can be seen in Table~\ref{tab:CompQY} that our final design in Fig.~\ref{QYdes}~(e) has the maximum gain bandwidth product. The proposed quasi-Yagi antenna has a modified ground structure for size reduction and offers a 22.4\% reduction in dipole length and 36\% reduction in the area of a single element antenna. Fig.~\ref{QYdes} shows the structure of the antenna at different design stages. The fabricated antenna and measurement setup is shown in Fig.~\ref{QYmes}. The measured results agree with the simulated results, and the fabrication tolerance can account for the slight discrepancies.

\section{Experimental Results and Analysis} \label{Experimental Results and Analysis}
For further performance evaluation of the designed Vivaldi and quasi-Yagi antennas, system level experiments were performed with the TWRoC. The experimental setup was placed in two different environments (A $\&$ B) and behind a brick wall, to test the two antennas. All the setups are real world scenarios where system level experiments were performed in the presence of clutter from static objects, multi-path propagation, etc. The obtained results are compared with that of a commercial off-the-shelf (COTS) pyramidal horn antenna. The horn antenna has a high directivity of 15~dBi but its large and bulky form-factor makes it unsuitable for practical applications using the TWRoC.

In all the conducted experiments, the TWRoC transmits an FMCW chirp between 2.052 and 2.6~GHz with a chirp rate of 11.2~MHz/$\mu$s at 0~dBm output power. The baseband signal from the radar output is fed to a Keysight 89601B vector signal analyzer (VSA), to obtain real time spectrograms. The VSA is set to acquire data with a frequency span of 1-20~MHz and samples the incoming IF signal with a bandwidth of 24.32~MHz.
\subsection{Spectrogram} \label{Spectrogram}
The short time fourier transform (STFT) is a standard method to study time-varying signals \cite{30749}. The STFT of a signal $x(t)$ is defined as 
\begin{align}
    X(t,w) =\frac{1}{\sqrt{2\pi}} \int_{-\infty}^{\infty} x(\tau)w^*(t-\tau) e^{-jw\tau} d\tau
\end{align}
where $w(t)$ is a suitably chosen analysis window.
The spectrogram corresponds to the squared magnitude of the STFT and is widely used in radar applications. The spectrogram can visualize the strength of the signals reflected from the targets over time at various frequencies \cite{330368}. 
For all the spectrogram plots in this paper, the x-axis represents frequencies between 1-5 MHz in steps of 5 kHz, and the resolution is 15~kHz with 12801 frequency points. The y-axis represents the time (in seconds) and the colour-map represents the power (\mbox{-66} to \mbox{-15}~dBm) of the received echoes. The power level of the colour-map is selected to minimize clutter from the environment and to make the signatures of interest clearly visible.
\subsection{Distance calculation from beat frequency} \label{maths}
Since we have used a long coaxial cable ($l_1$~=40~m, $\epsilon _{r1}$) and other RF cables ($l_2$~=7.45~m, $\epsilon _{r2}$), the target is no longer located at a distance of $d$ from the radar. We need to modify (\ref{eq4}) to compensate for the additional length which will be reflected in the observed beat frequencies. The total distance of the target from the radar is $D$ given by:
\begin{align}
    D = 2d + l_1 + l_2
\end{align}
Incorporating the different lengths and dielectric constants of the cables, we modify (\ref{eq4}) as:
\begin{equation} \label{eq9} 
    f_b = s \left[ \frac{2d}{c} + \frac{l_1}{v_1} +\frac{l_2}{v_2} \right] 
\end{equation} 
where, $v_1$ and $v_2$ are the velocities of wave propagation inside the cables with lengths $l_1$ and $l_2$. Hence,
\begin{equation} \label{eq10} 
    d = \frac{s}{c} \left[ 2d + d_0 \right]  
\end{equation}
where $d_0$ is a constant representing the effective length of both the cables given by $d_0 =l_1\epsilon_{r1} + l_2\epsilon_{r2}$.
\begin{equation} \label{calc} 
    d = \frac{1}{2}\left( \frac{cf_b}{s} - d_0 \right)  
\end{equation}
We have used the modified equation (\ref{eq10}) for all our calculations instead of (\ref{eq4}). Once we know the value of the constant $d_0$, we can calculate the distance $d$ of the target from the antennas using (\ref{calc}). The constant $d_0$ for both the environments as well as for the through-wall setup is 55.97~m. 

\subsection{Environment A} \label{Environment A}
\begin{figure*}[ht]
\centering
\subfloat[]{\includegraphics[width=0.25\textwidth,keepaspectratio]{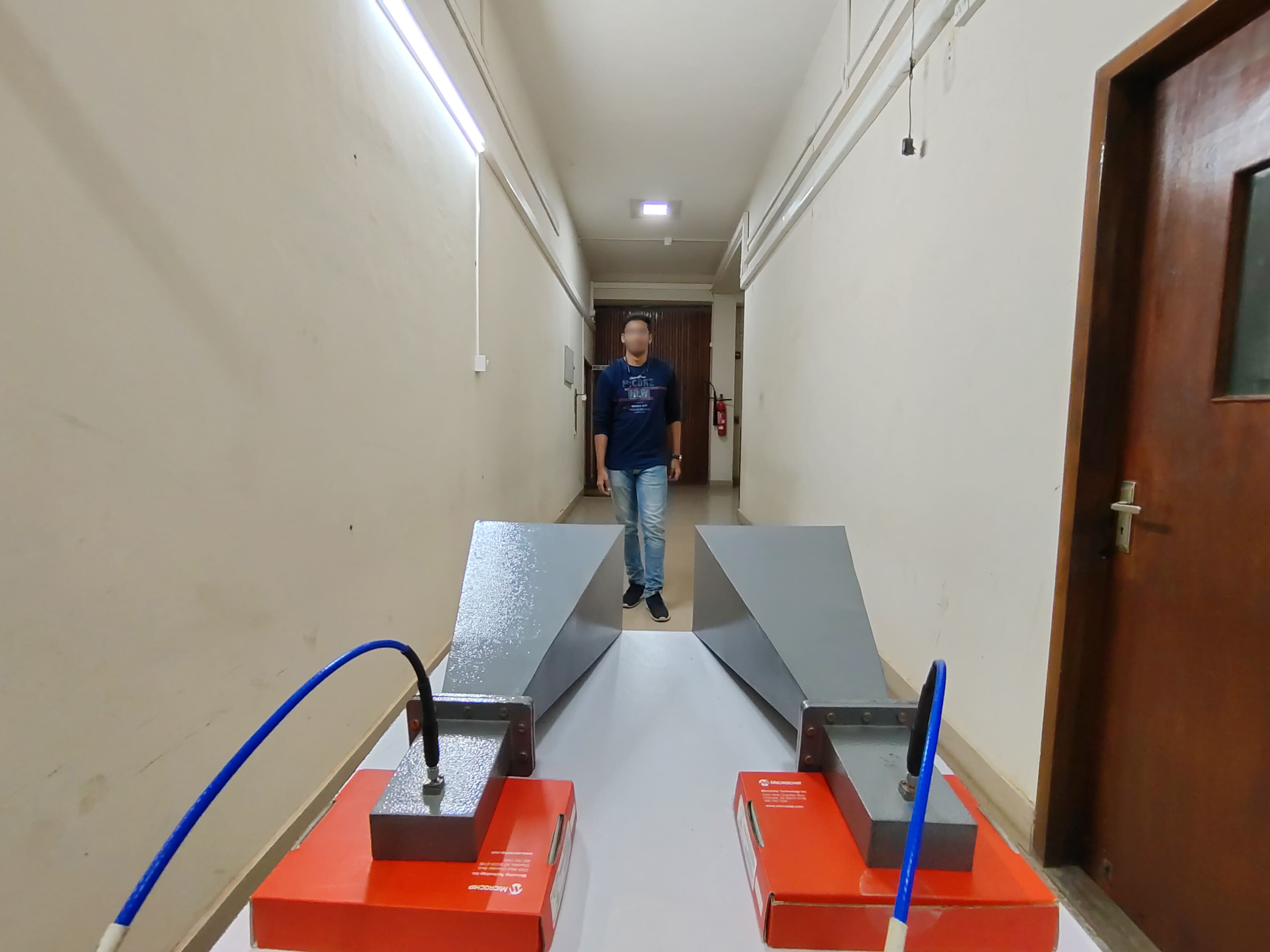}%
\label{1a}}
\hfil
\subfloat[]{\includegraphics[width=0.25\textwidth,keepaspectratio]{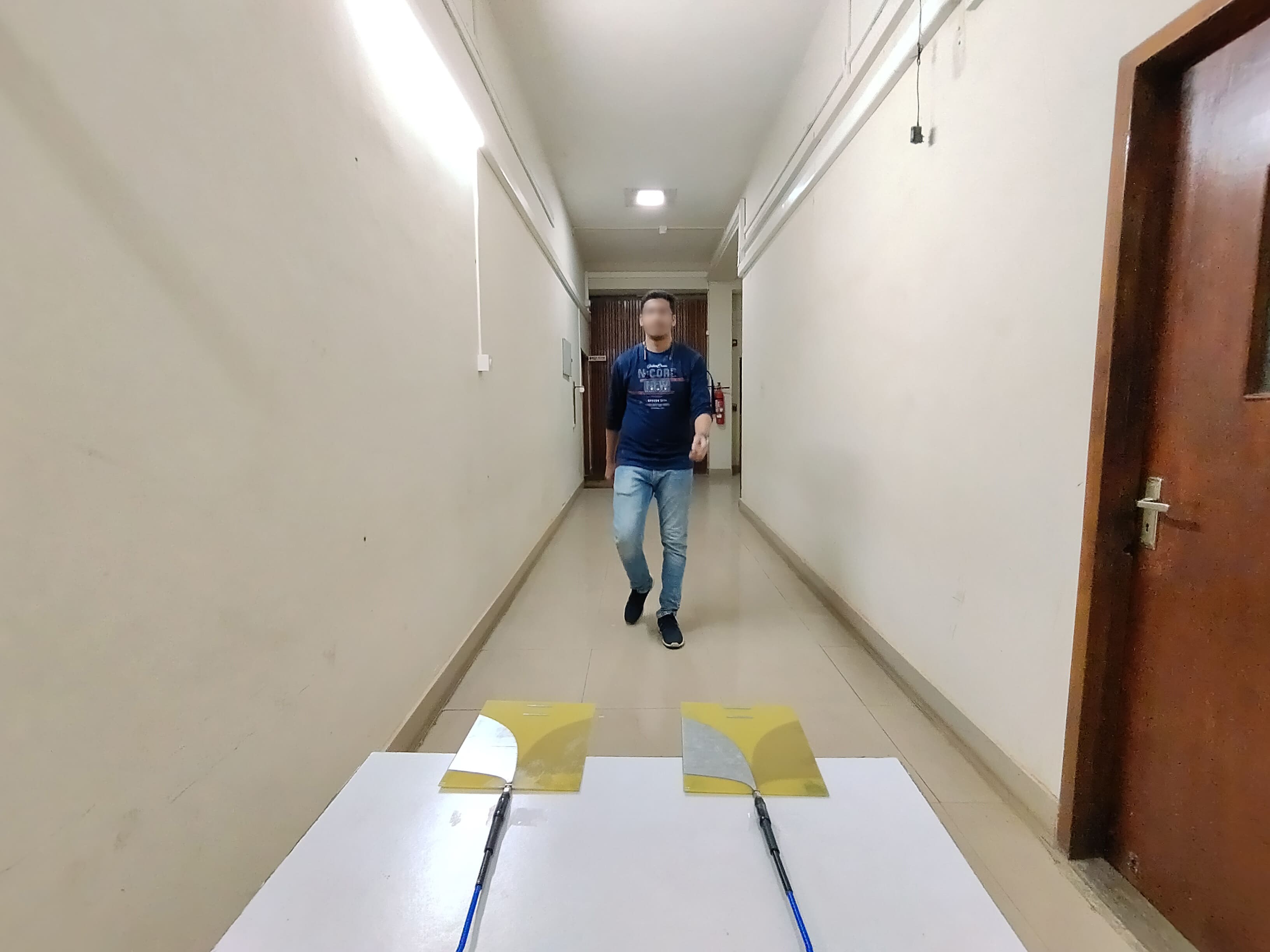}%
\label{1b}}
\hfil
\subfloat[]{\includegraphics[width=0.25\textwidth,keepaspectratio]{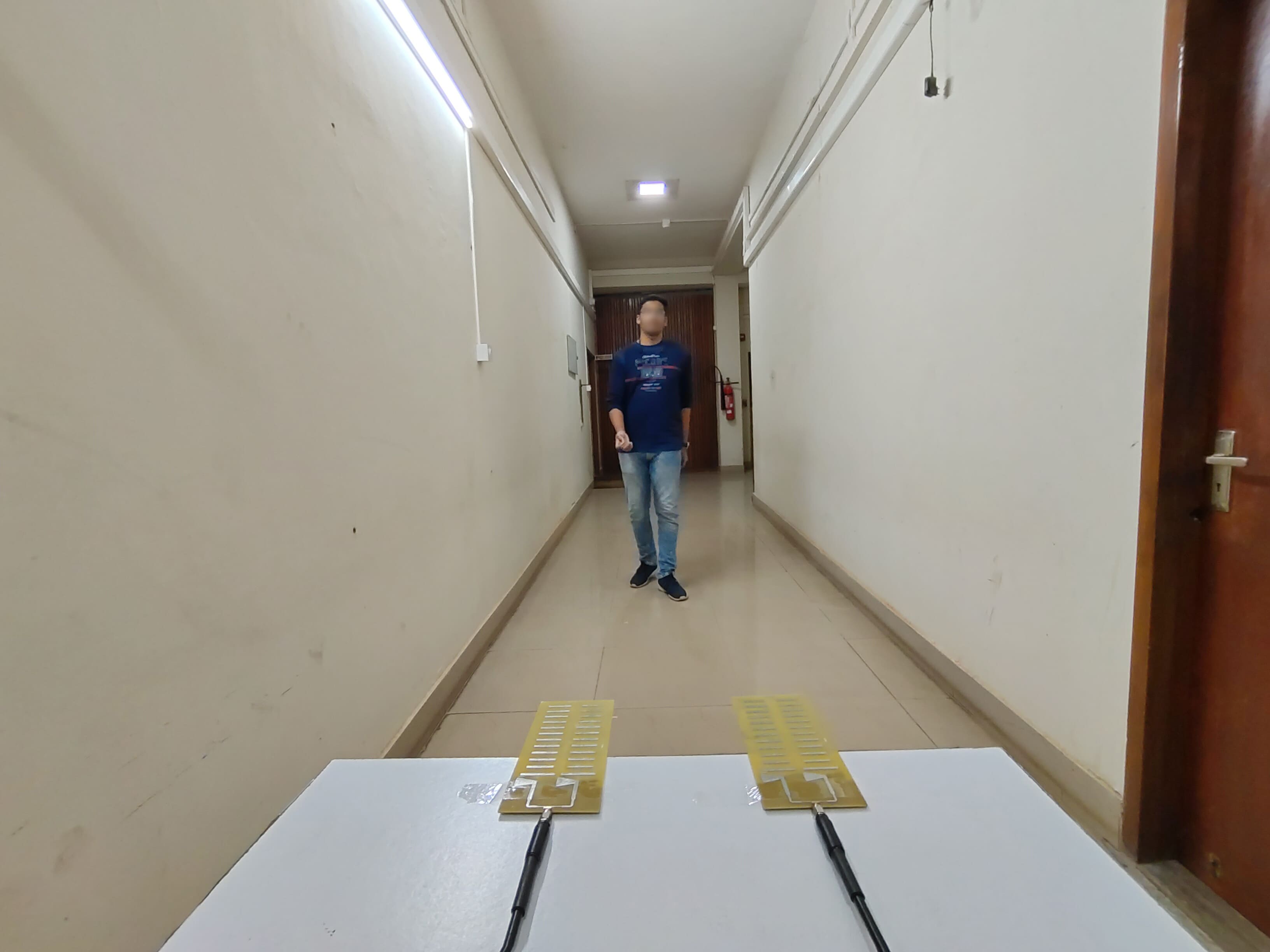}%
\label{1c}} \vspace{-0.4cm}
\hfil
\subfloat[]{\includegraphics[width=0.32\textwidth]{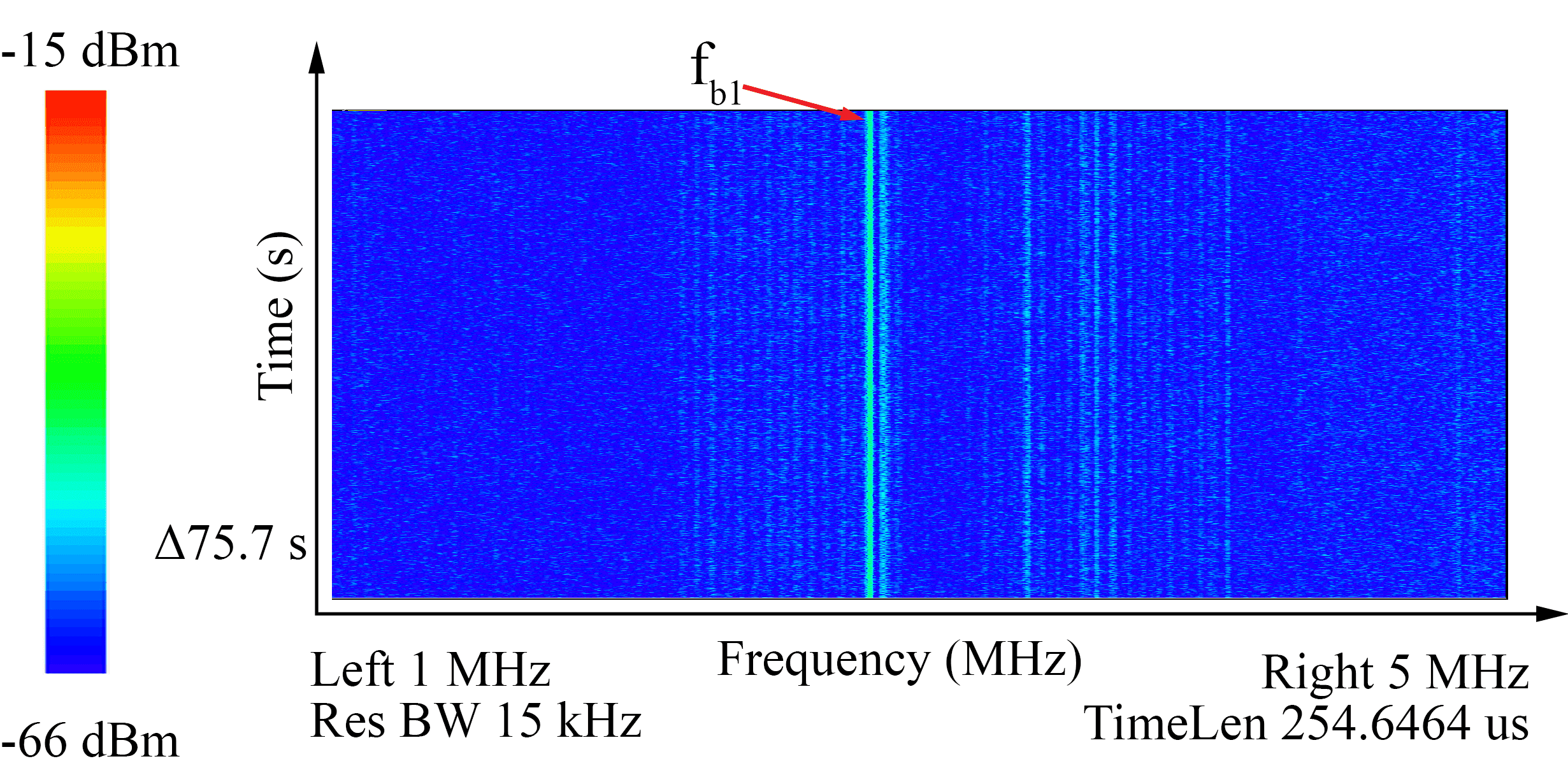}%
\label{1d}}
\hfil
\subfloat[]{\includegraphics[width=0.32\textwidth]{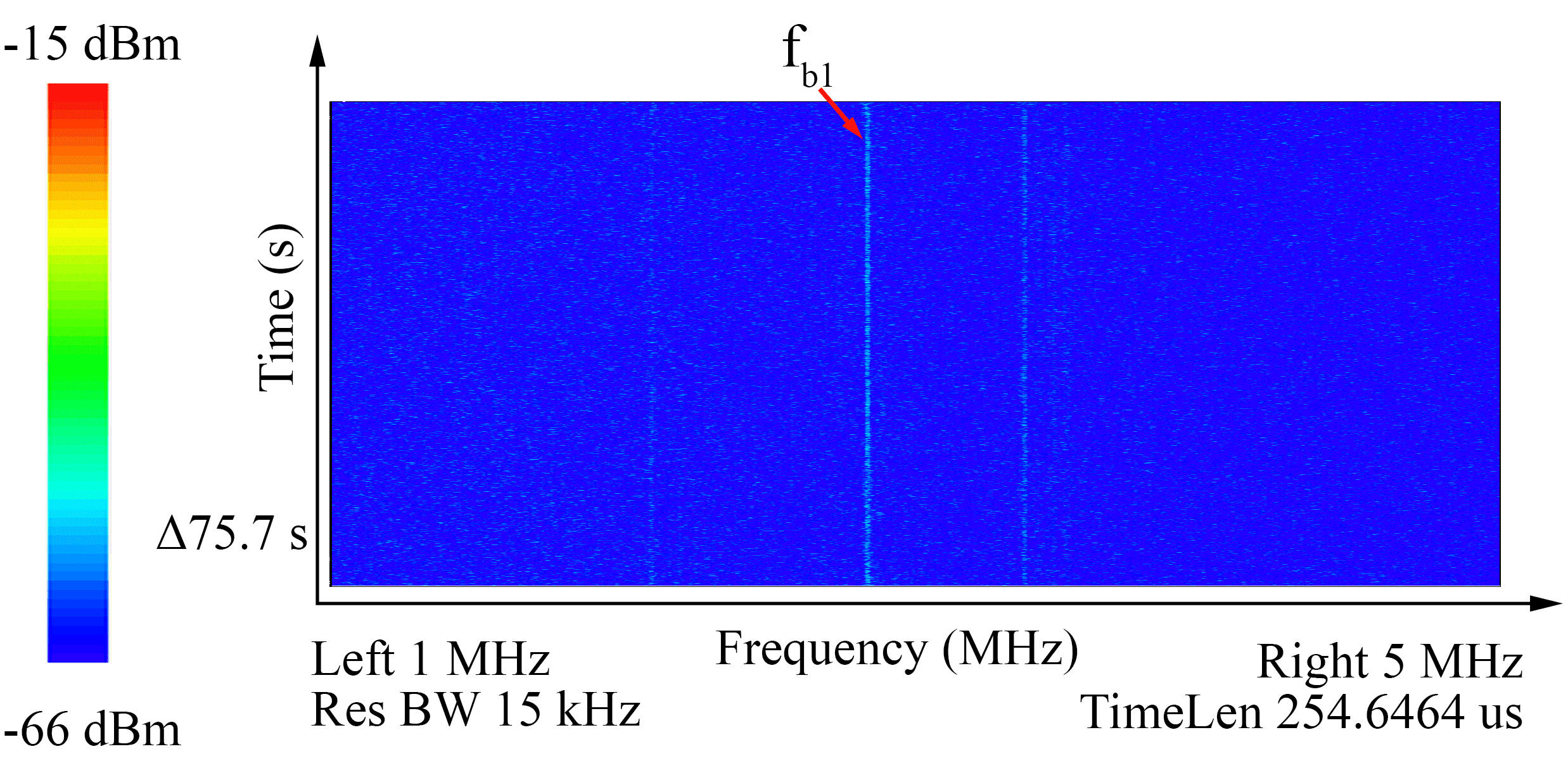}%
\label{1e}}
\hfil
\subfloat[]{\includegraphics[width=0.32\textwidth]{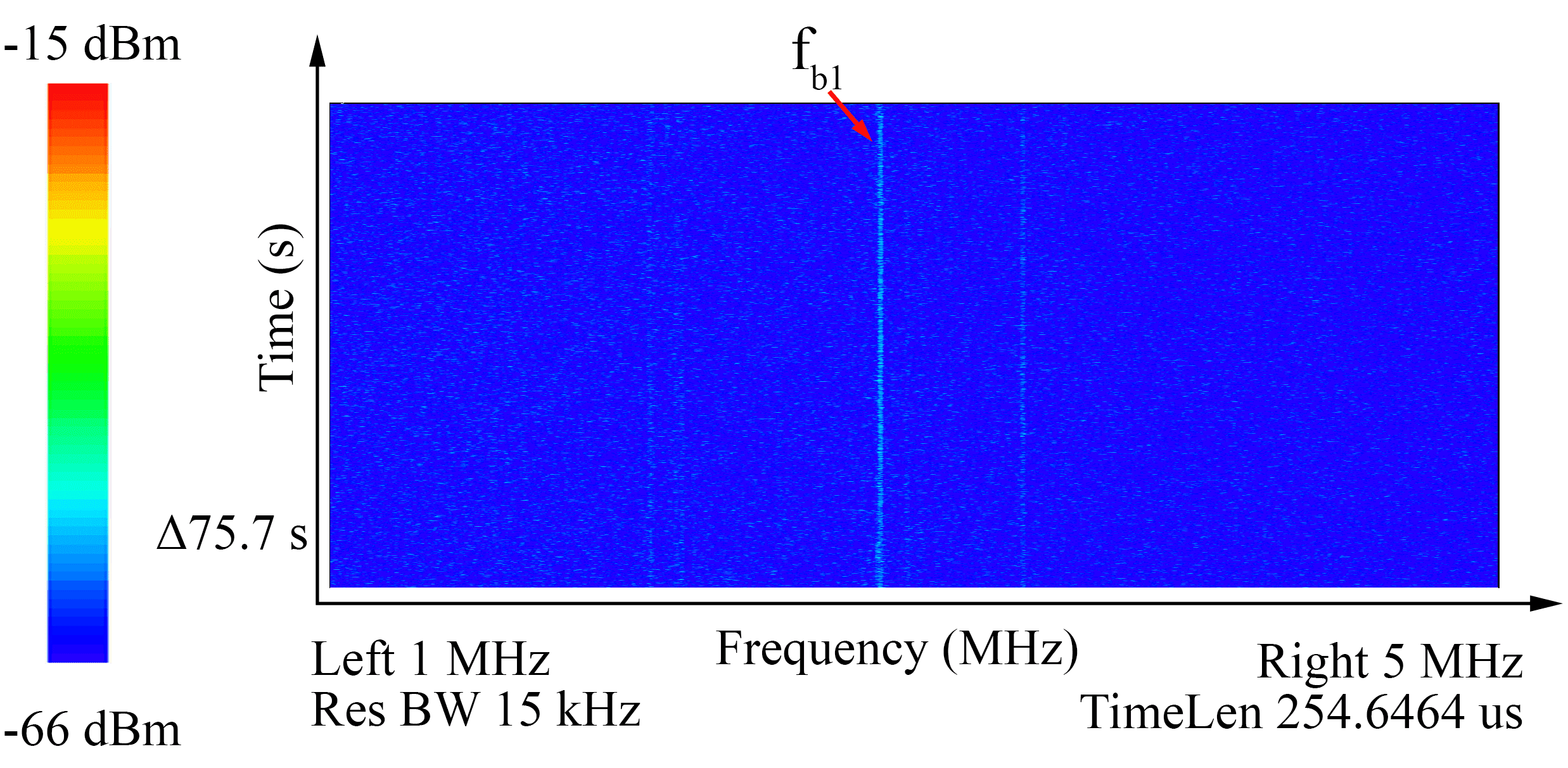}%
\label{1f}} \vspace{-0.4cm}
\hfil
\subfloat[]{\includegraphics[width=0.32\textwidth]{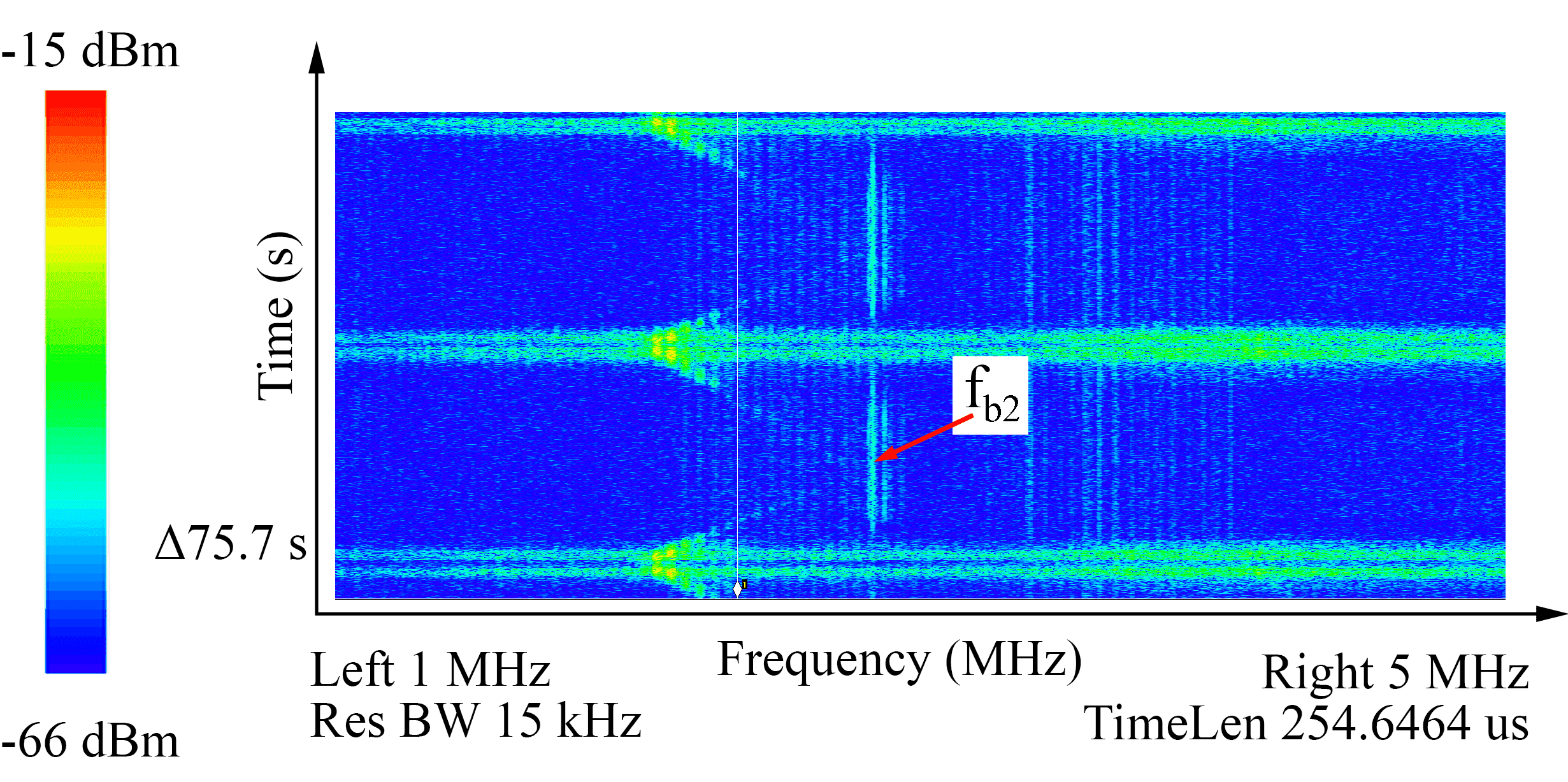}%
\label{1g}}
\hfil
\subfloat[]{\includegraphics[width=0.32\textwidth]{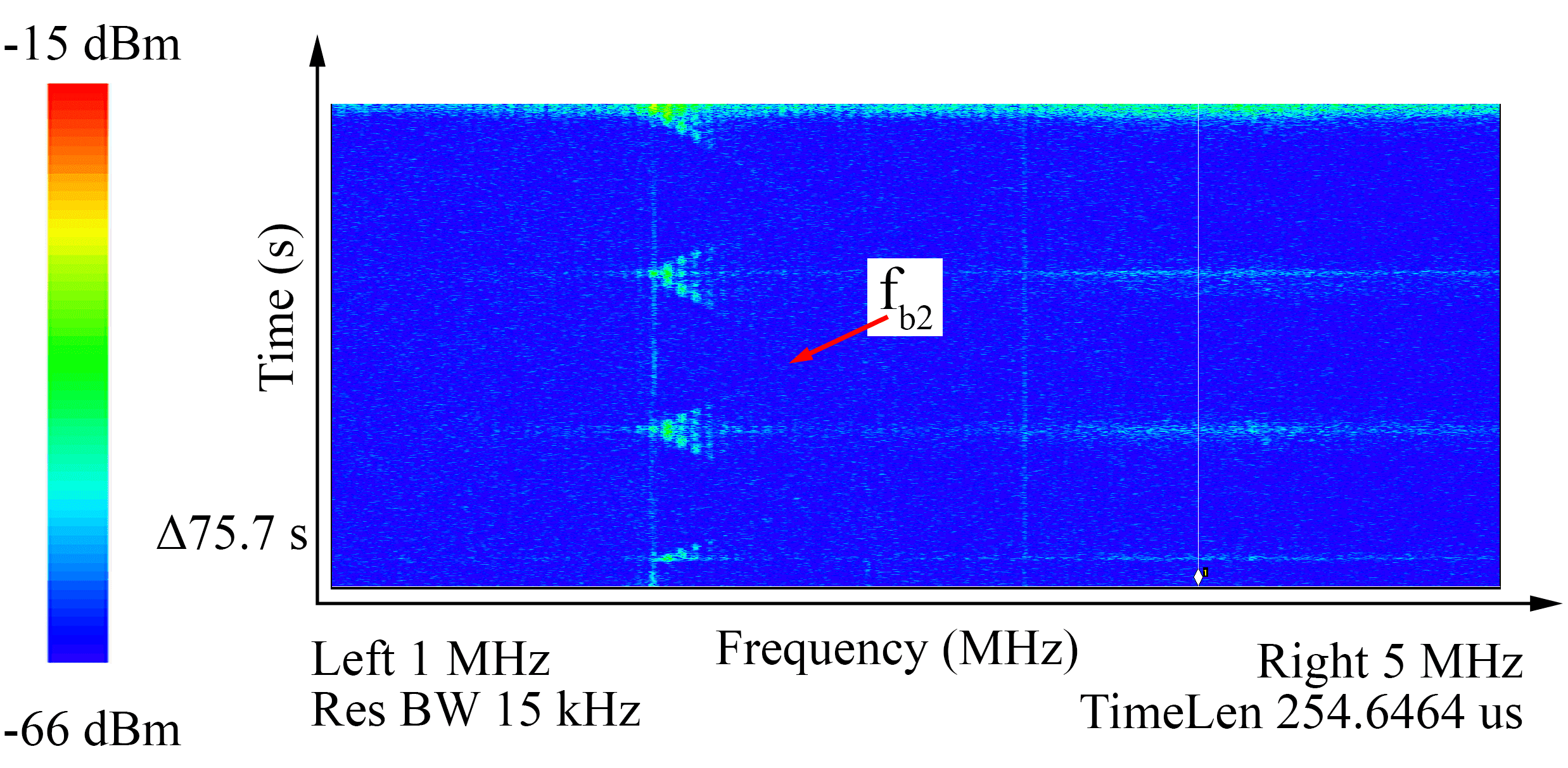}%
\label{1h}}
\hfil
\subfloat[]{\includegraphics[width=0.32\textwidth]{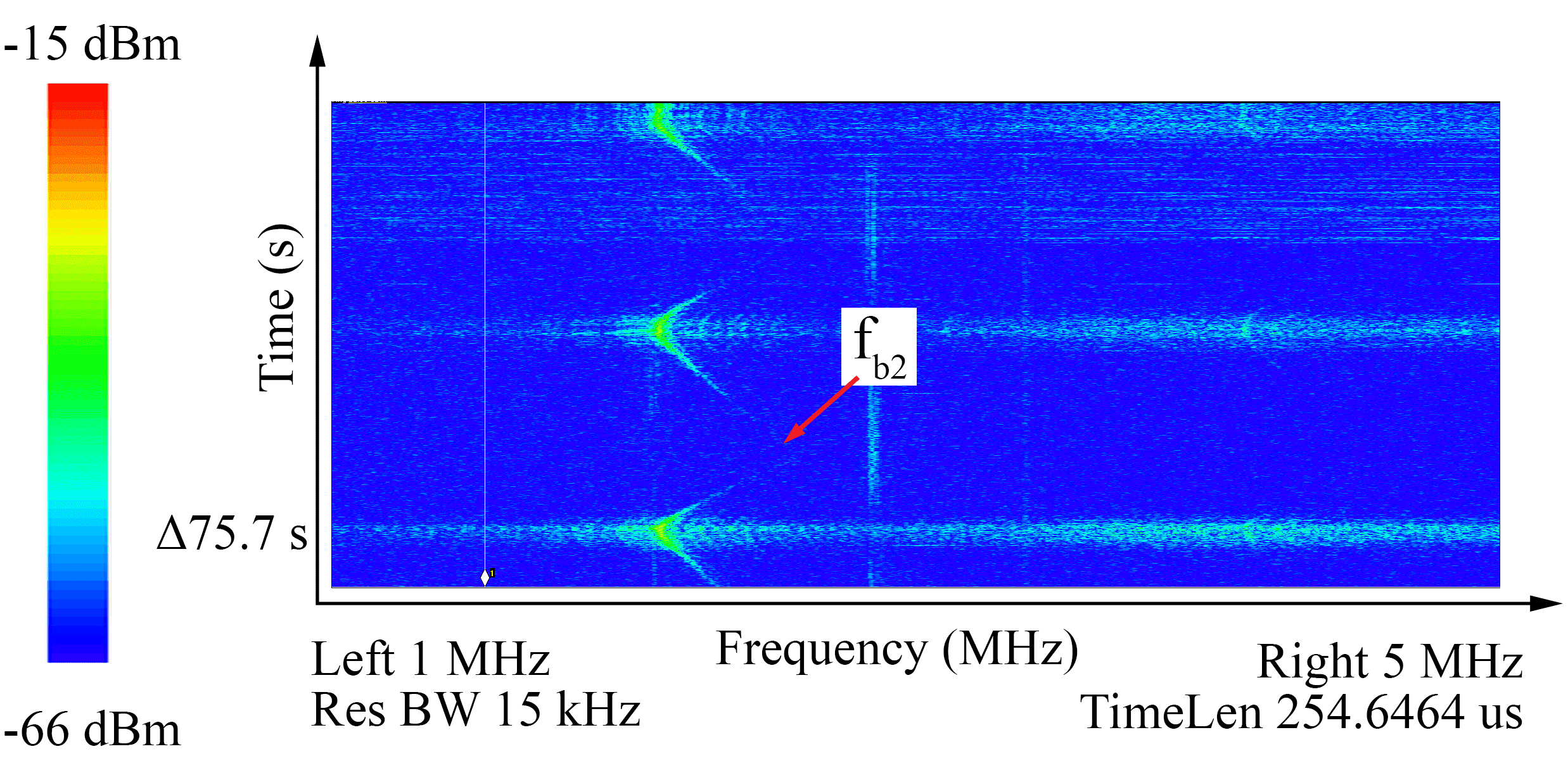}%
\label{1i}}

\caption{Snapshots showing the experimental setup (a), (b), (c), spectrograms of the static environment (d), (e), (f) and spectrograms of the walking human (g), (h), (i) in environment A for the horn, Vivaldi and quasi-Yagi antennas respectively.}
\label{corridor}
\vspace{-0.6cm}
\end{figure*}

Environment A (Fig. \ref{corridor}) is a corridor with a wall at one end, with the antennas placed at a distance of 10~m from the wall. The spectrograms for the static environment with no moving targets, obtained with  the horn, Vivaldi and quasi-Yagi antennas, is shown in Figs. \ref{corridor}~(d), (e) and (f) respectively. We expect the spectrogram to have a prominent straight line at the beat frequency corresponding to the distance of the brick wall from the antenna. Table \ref{tab:Environment A} lists the observed beat frequency ($f_{b_1}$) and the corresponding distance of the wall ($d_{obs_1}$) for all three antennas. In addition, several weaker lines are observed, because of multiple reflections due to a scattering-rich environment. A volunteer then walks back and forth between the antenna and the brick wall, and the snapshots of the corresponding spectrograms are shown in Figs. \ref{corridor}~(g), (h) and (i), for the horn, Vivaldi and quasi-Yagi antennas, respectively. We observe that the spectrogram has a triangular waveform depicting the back and forth movement of the person. In Table \ref{tab:Environment A}, we have listed the maximum beat frequency ($f_{b_2}$) and the corresponding distance ($d_{obs_2}$) at which the moving target is identified by the same amount of received power in the spectrogram, representing the sensitivity of the setup. Clearly, the horn antenna is able to look the furthest, but the performance of the quasi-Yagi antenna is quite comparable.

\begin{table}
\centering
\caption{\textsc{Environment A}}
\label{tab:Environment A}
\resizebox{0.3\columnwidth}{!}{%
\begin{tabular}{@{}ccccc@{}}
\toprule
\textbf{}  & \multicolumn{2}{c}{\textbf{Static}} & \multicolumn{2}{c}{\textbf{Person Walking}} \\ \midrule
\textbf{Antenna} &
  \textbf{\begin{tabular}[c]{@{}c@{}}$f_{b_1}$  \\ (MHz)\end{tabular}} &
  \textbf{\begin{tabular}[c]{@{}c@{}}$d_{obs_1}$ \\ (m)\end{tabular}} &
  \textbf{\begin{tabular}[c]{@{}c@{}}$f_{b_2}$\\ (MHz)\end{tabular}} &
  \textbf{\begin{tabular}[c]{@{}c@{}}$d_{obs_2}$ \\ (m)\end{tabular}} \\ \midrule
Horn       & 2.908            & 10.32            & 2.703                 & 7.42                \\
Vivaldi    & 2.837            & 10.32            & 2.477                 & 5.47                \\
Quasi-Yagi & 2.836            & 10.32            & 2.563                 & 6.64                \\ \bottomrule
\end{tabular}%
} \vspace{-0.4cm}
\end{table}
\subsection{Environment B} \label{Environment B}
\begin{figure*}
\centering
\subfloat[]{\includegraphics[width=0.25\textwidth]{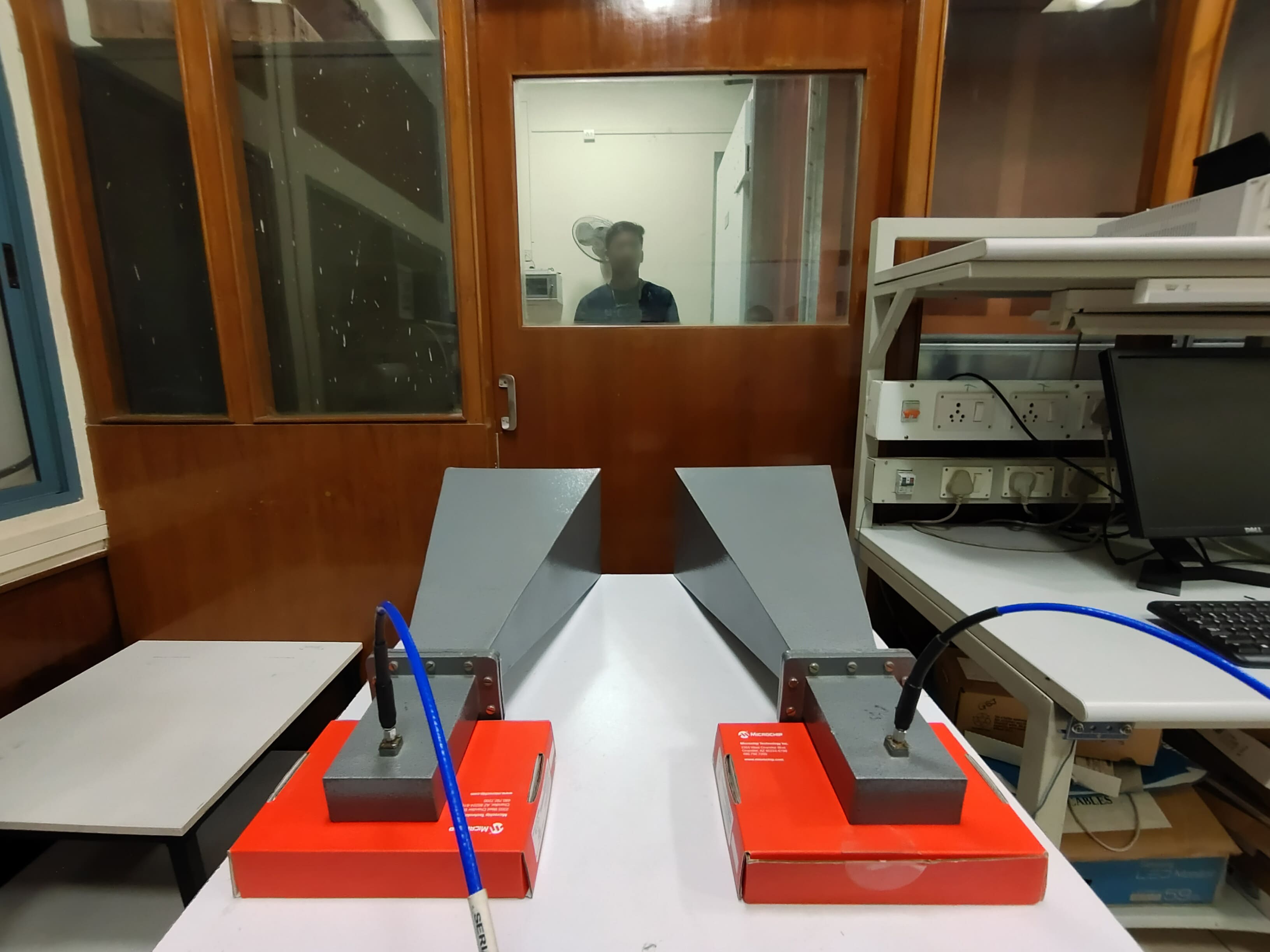}%
\label{3a}}
\hfil
\subfloat[]{\includegraphics[width=0.25\textwidth]{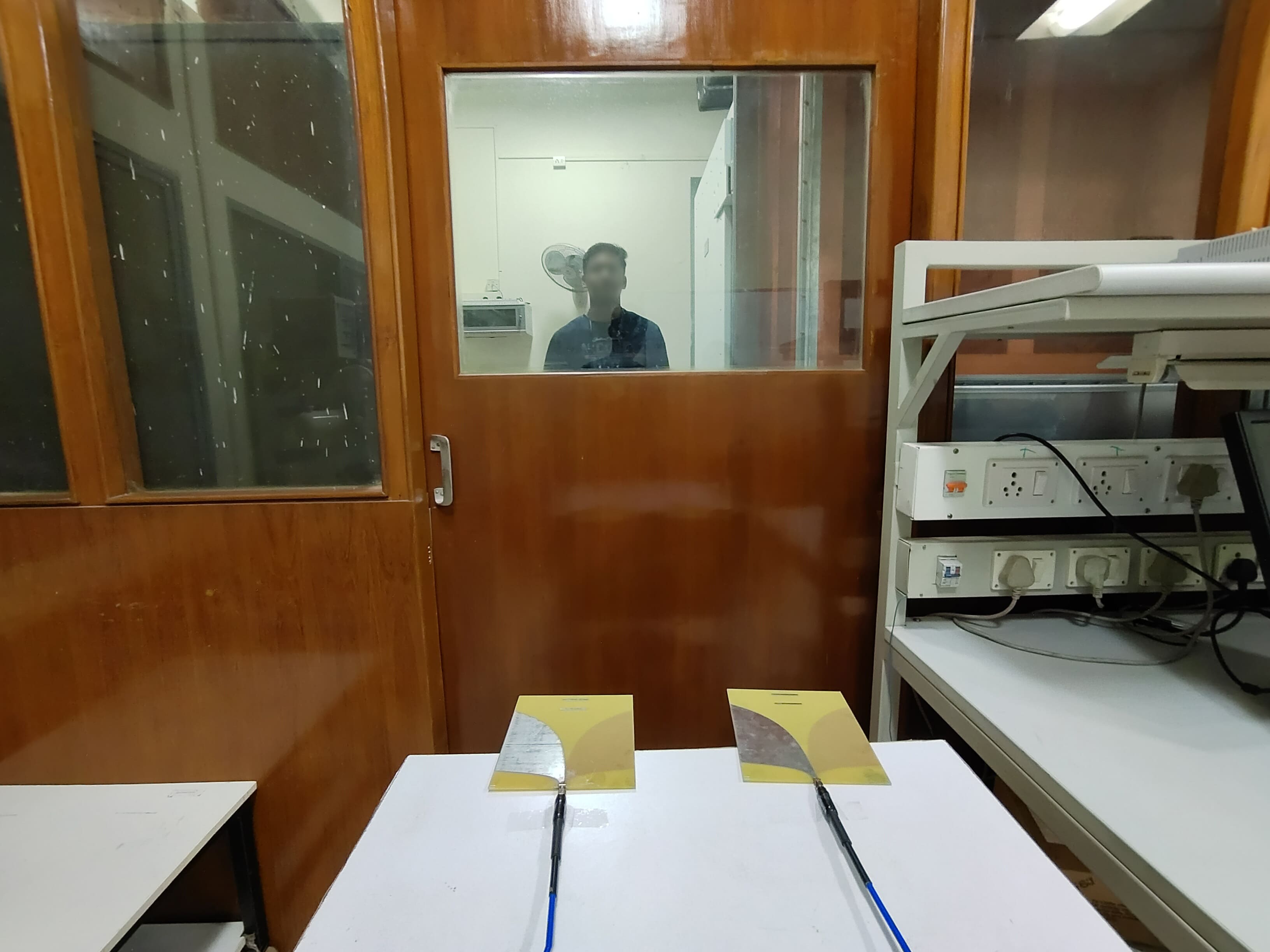}%
\label{3b}}
\hfil
\subfloat[]{\includegraphics[width=0.25\textwidth]{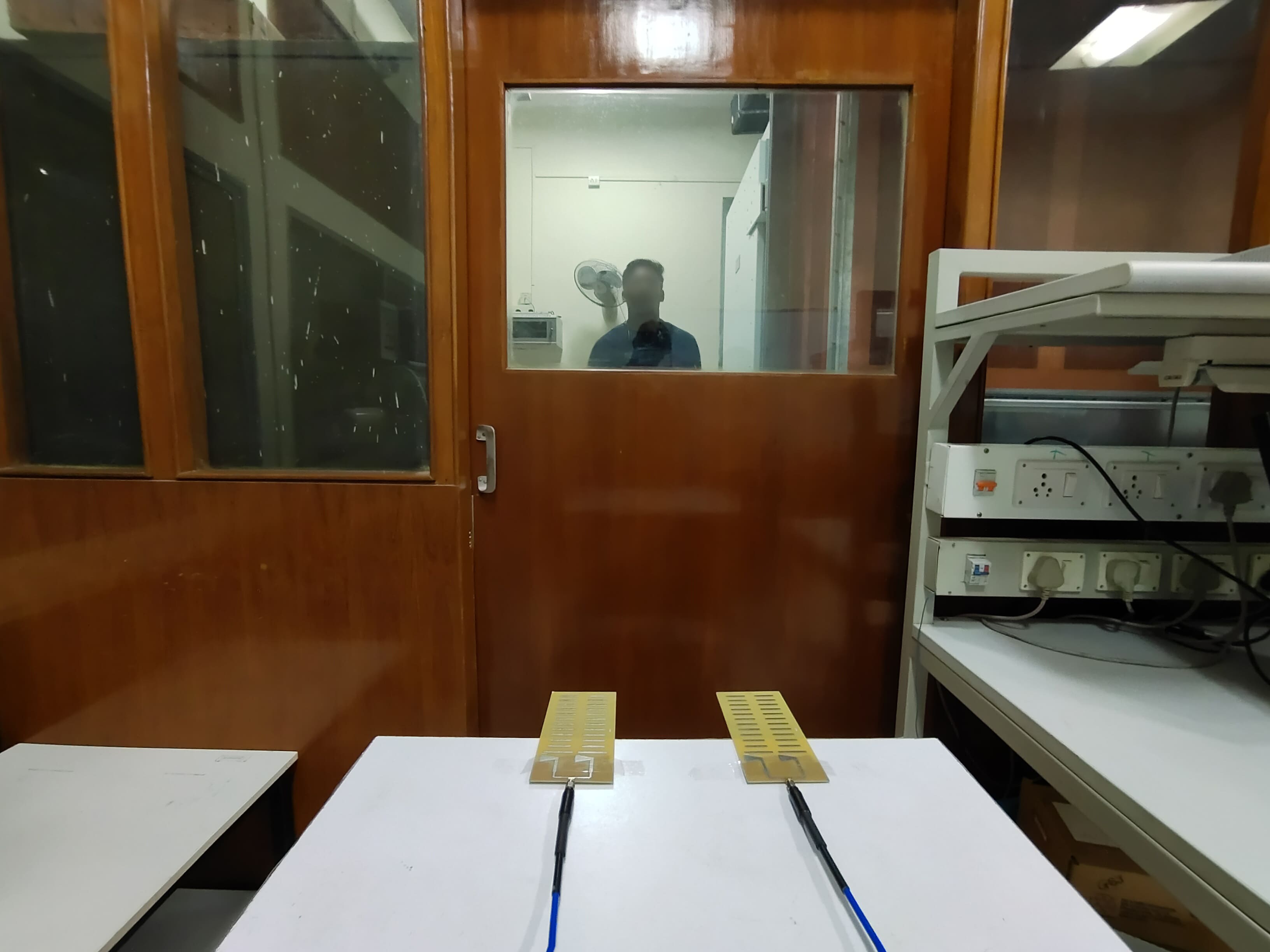}%
\label{3c}} \vspace{-0.4cm}
\hfil
\subfloat[]{\includegraphics[width=0.32\textwidth]{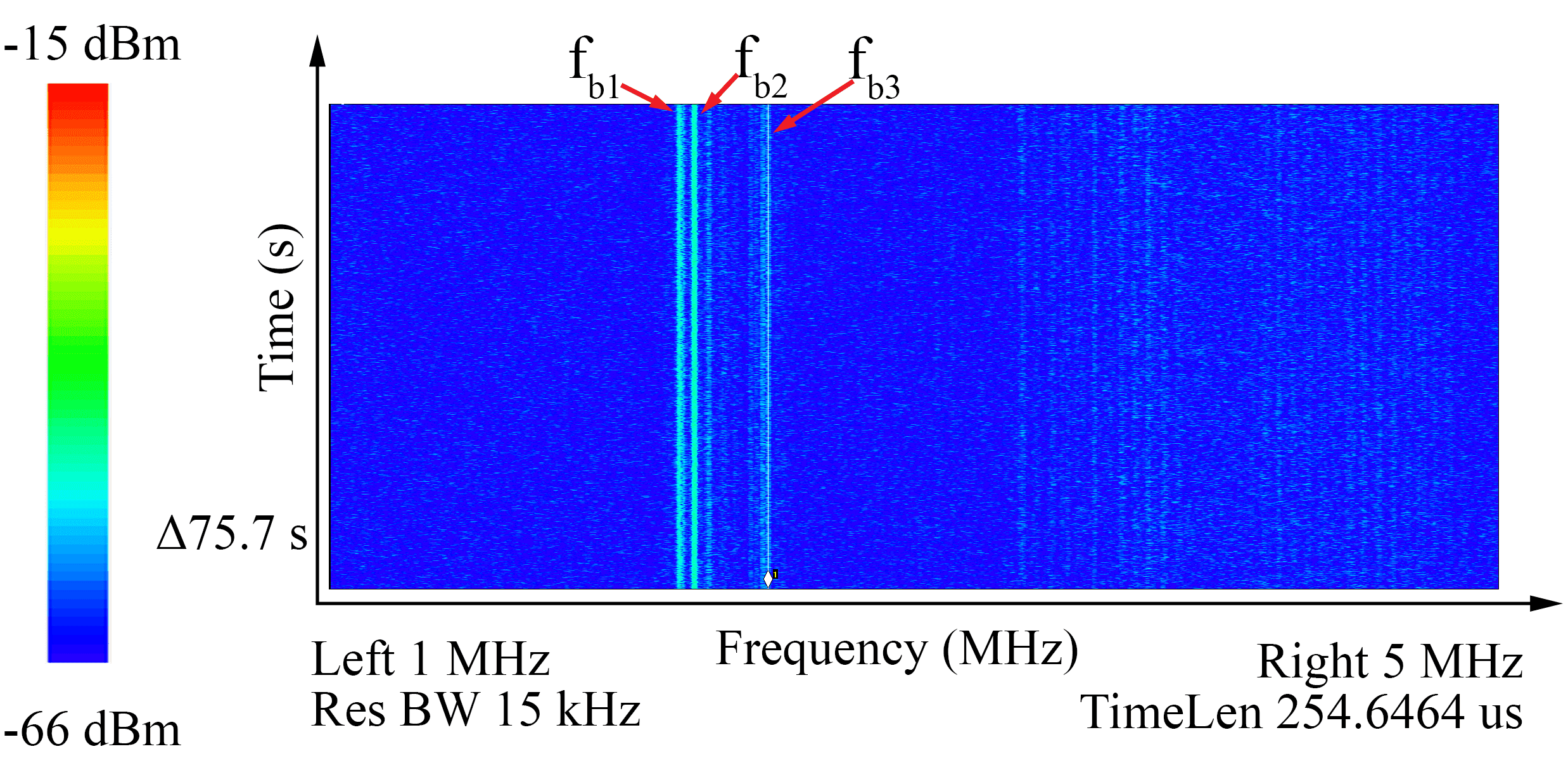}%
\label{3d}}
\hfil
\subfloat[]{\includegraphics[width=0.32\textwidth]{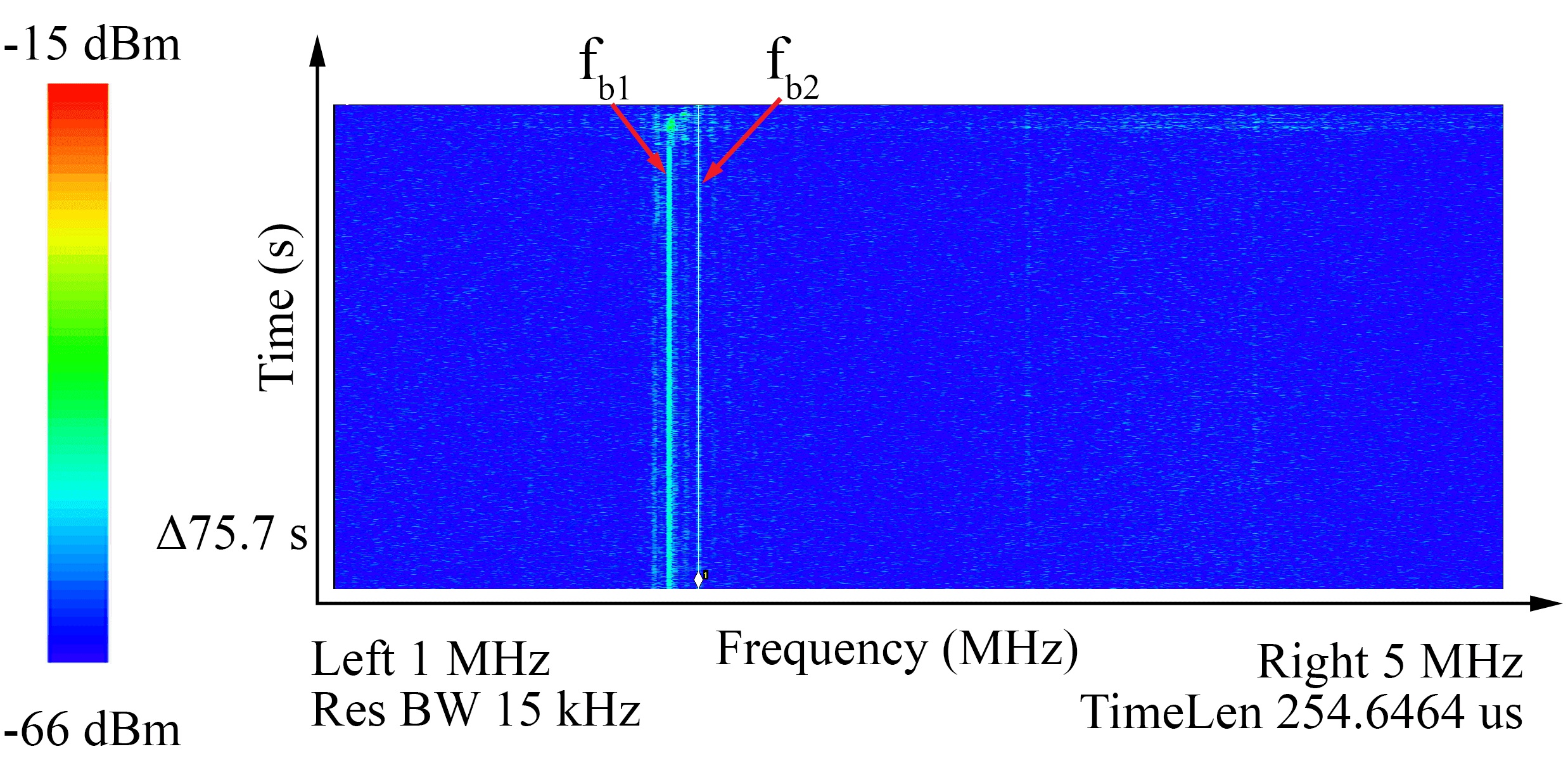}%
\label{3e}}
\hfil
\subfloat[]{\includegraphics[width=0.32\textwidth]{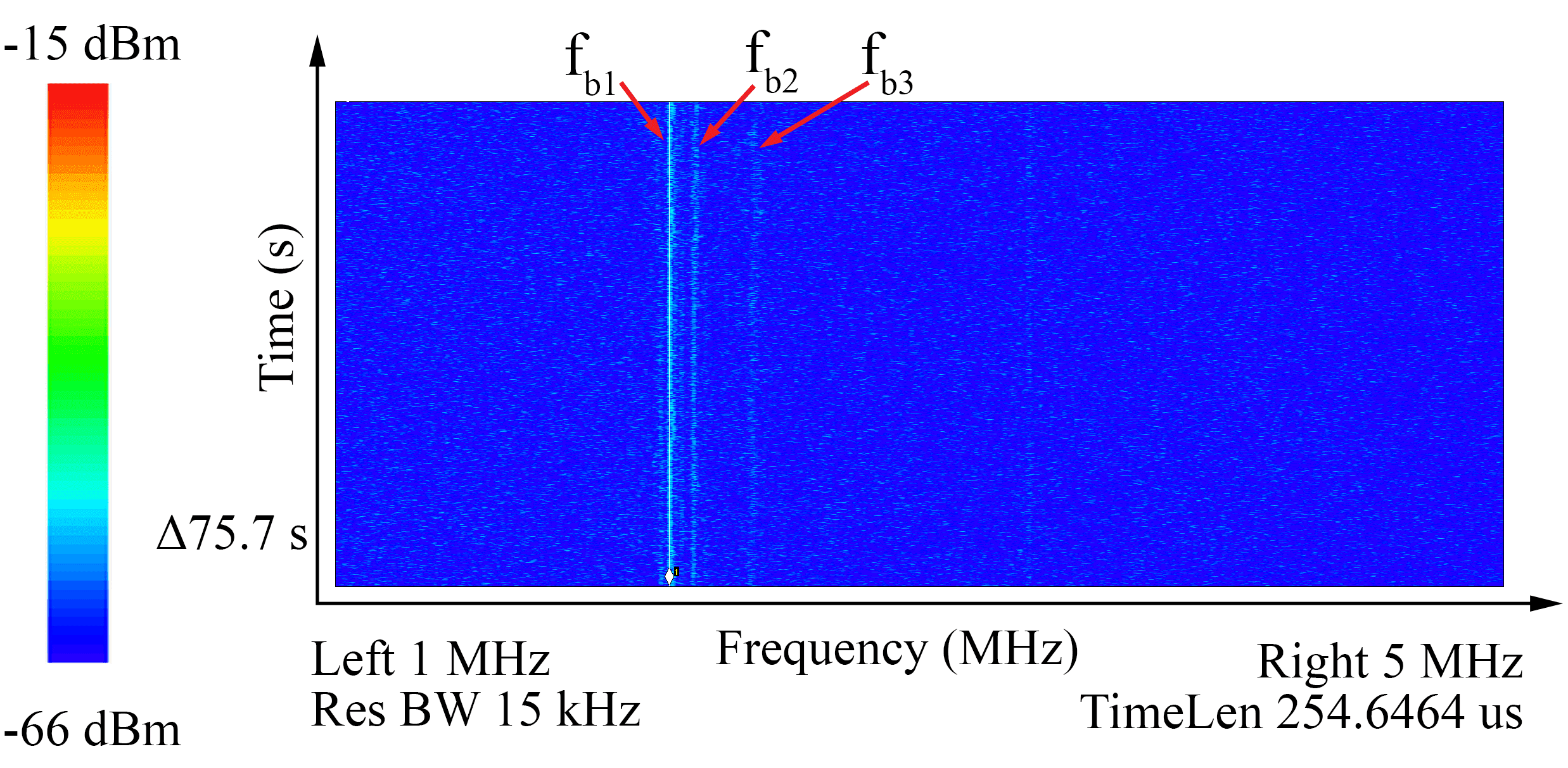}%
\label{3f}} \vspace{-0.4cm}
\hfil
\subfloat[]{\includegraphics[width=0.32\textwidth]{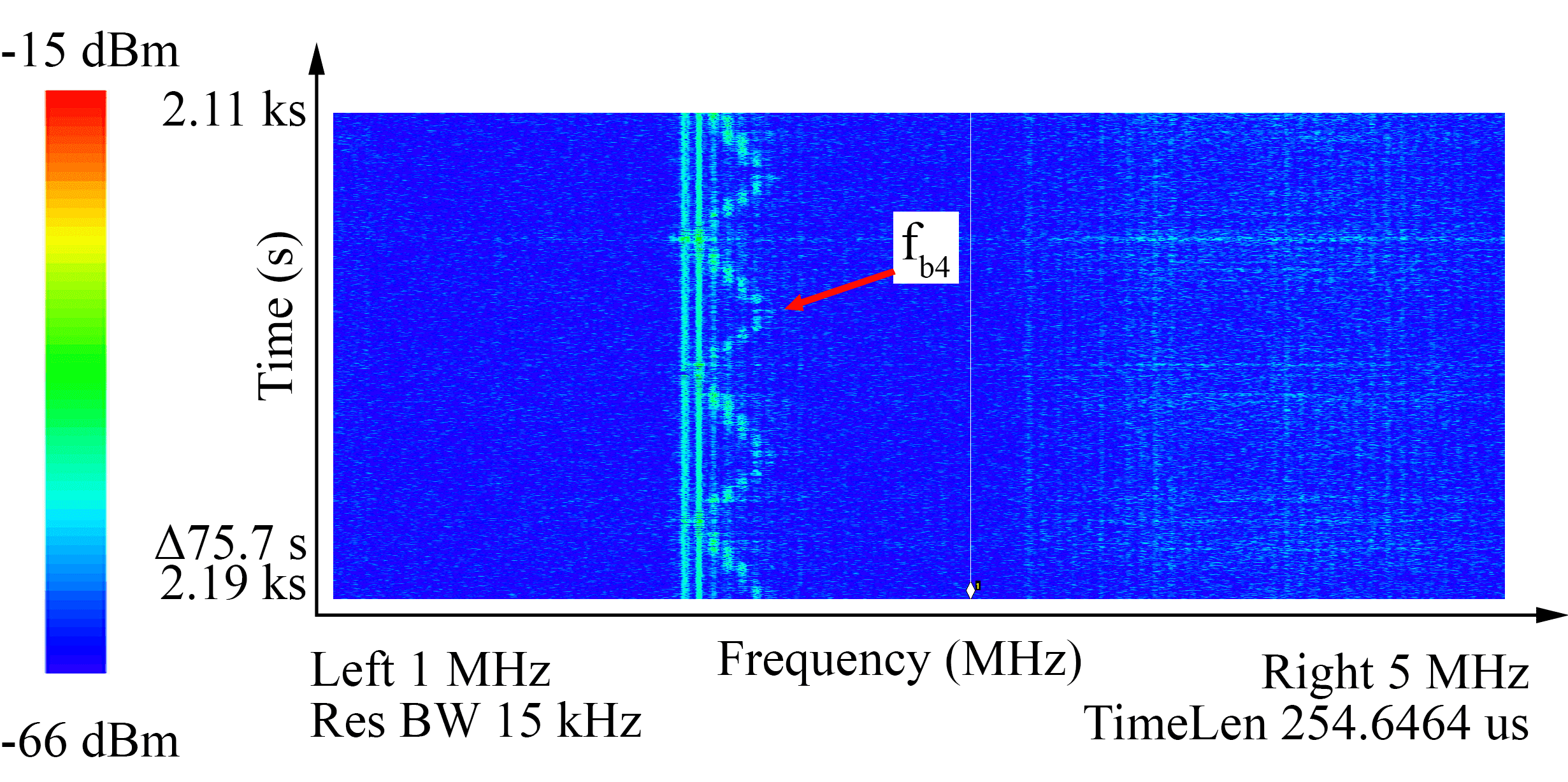}%
\label{3g}}
\hfil
\subfloat[]{\includegraphics[width=0.32\textwidth]{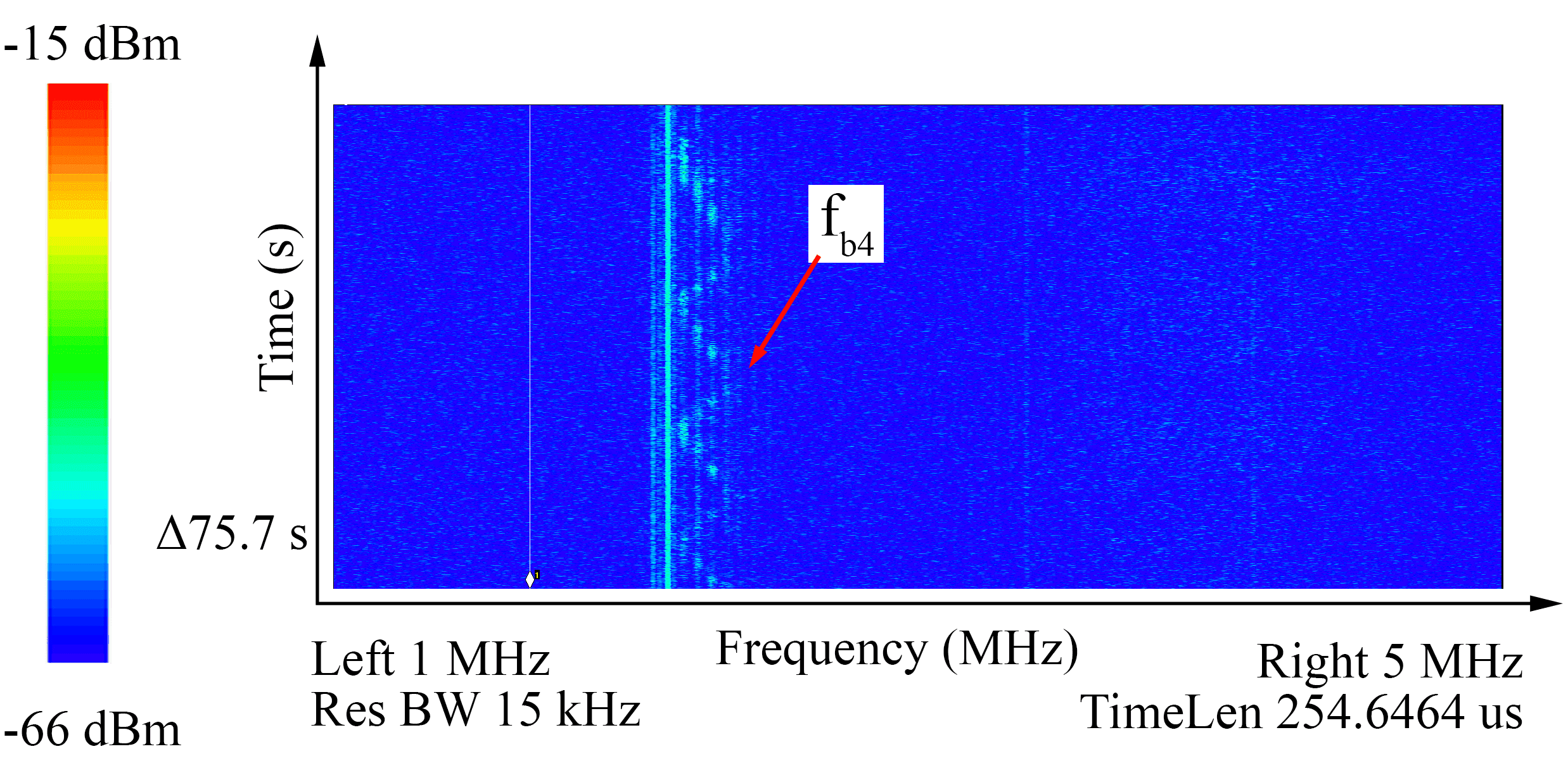}%
\label{3h}}
\hfil
\subfloat[]{\includegraphics[width=0.32\textwidth]{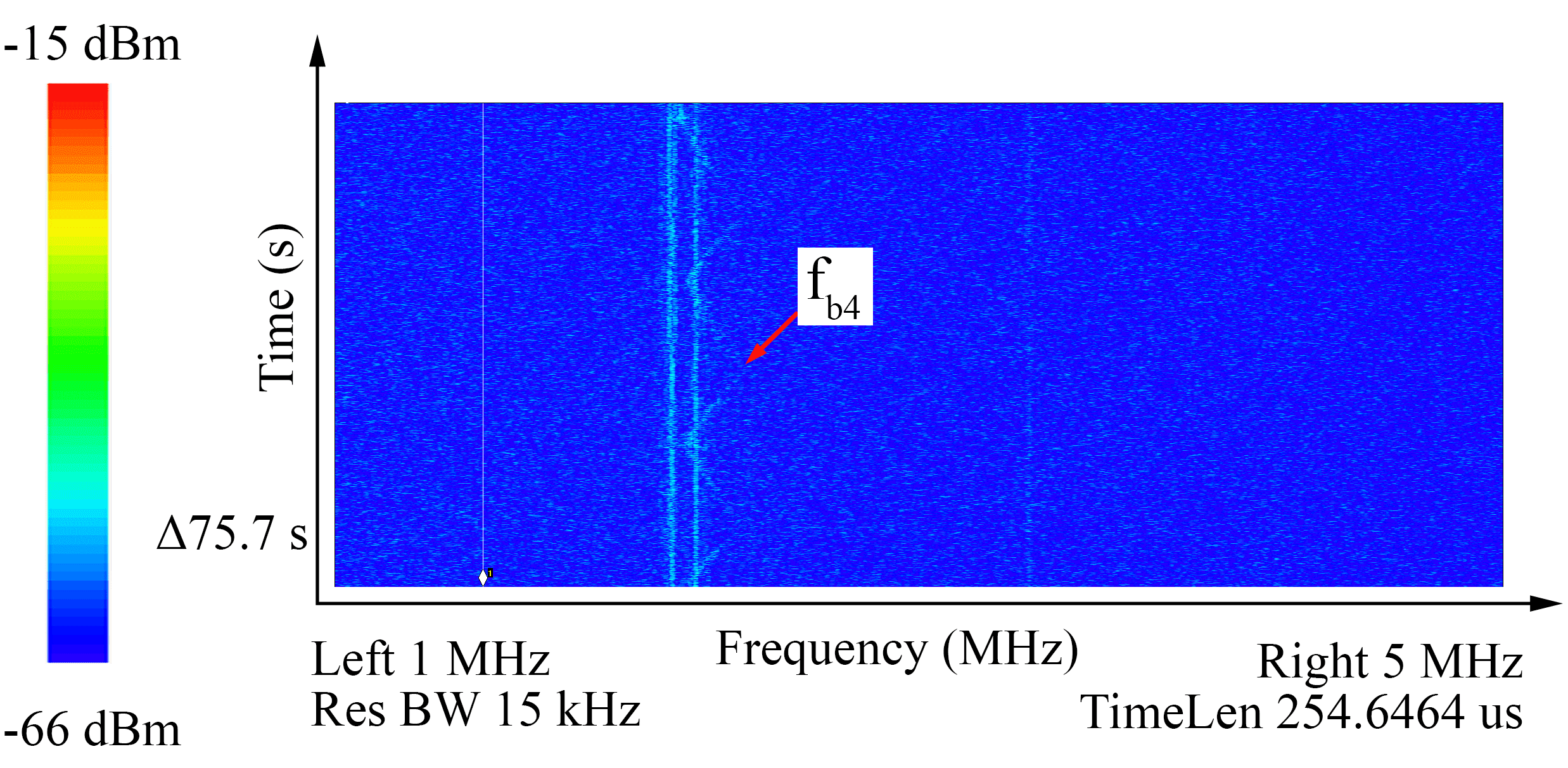}%
\label{3i}}
	\caption{Snapshots showing the experimental setup (a), (b), (c), spectrograms of the static environment (d), (e), (f) and spectrograms of the walking human (g), (h), (i) with the door closed for the horn, Vivaldi and quasi-Yagi antennas respectively.}
	\label{closedoor}
\vspace{-0.4cm}
\end{figure*}

In environment B, a section of our laboratory is used. It has a wooden partition, with a sliding wooden door that acts like a low-loss wall in our experiments. The antennas are placed at a distance of 4.5~m from the back wall and the wooden partition lies at a distance of 2~m from the antennas. A volunteer walked back and forth in front of the antennas while the wooden door was kept 
closed.
The experimental setup, using the closed door, is shown in Figs. \ref{closedoor}~(a), (b) and (c). The wooden door was closed to to create a "wall-like" scenario. The volunteer then walked back and forth, behind the wooden door. First, the spectrograms for a static environment were obtained, as shown in Figs. \ref{closedoor}~(d), (e) and (f).
Table~\ref{tab:Environment B: Closed Door} lists the observed beat frequencies ($f_{b_{1}},f_{b_{2}},f_{b_{3}}$) and the corresponding distances ($d_{obs_{1}},d_{obs_{2}},d_{obs_{3}}$) at which the static targets were identified by the three antennas. 
The closed door provides enough attenuation to make the reflections from the back wall less prominent. 
In the process, additional reflections due to a nearby metallic reflector are now visible. The measured distance of the wooden partition, metallic reflector and the back wall from the antennas were $d'_1$ = 1~m, $d'_2$ = 2.5~m and $d'_3$ = 4.5~m, respectively.
Next, the spectrograms for the volunteer walking behind the wooden door, were obtained, as shown in Figs. \ref{closedoor}~(g), (h) and (i). Table~\ref{tab:Environment B: Closed Door} lists the maximum beat frequency ($f_{b_4}$) and the corresponding distance ($d_{obs_4}$) up to which the signatures of the volunteer can be traced.
\begin{table*}[ht]
\centering
\caption{\textsc{Environment B: Closed Door}}
\label{tab:Environment B: Closed Door}
\resizebox{0.6\textwidth}{!}{%
\begin{tabular}{@{}ccccccccc@{}}
\toprule
\textbf{}  & \multicolumn{6}{c}{\textbf{Static}}        & \multicolumn{2}{c}{\textbf{Person Walking}} \\ \midrule
\textbf{Antenna} &
  \textbf{\begin{tabular}[c]{@{}c@{}}$f_{b_{1}}$  \\ (MHz)\end{tabular}} &
  \textbf{\begin{tabular}[c]{@{}c@{}}$d_{obs_1}$ \\ (m)\end{tabular}} &
  \textbf{\begin{tabular}[c]{@{}c@{}}$f_{b_{2}}$  \\ (MHz)\end{tabular}} &
  \textbf{\begin{tabular}[c]{@{}c@{}}$d_{obs_2}$ \\ (m)\end{tabular}} &
  \textbf{\begin{tabular}[c]{@{}c@{}}$f_{b_{3}}$  \\ (MHz)\end{tabular}} &
  \textbf{\begin{tabular}[c]{@{}c@{}}$d_{obs_3}$ \\ (m)\end{tabular}} &
  \textbf{\begin{tabular}[c]{@{}c@{}}$f_{b_{4}}$  \\ (MHz)\end{tabular}} &
  \textbf{\begin{tabular}[c]{@{}c@{}}$d_{obs_4}$ \\ (m)\end{tabular}} \\ \midrule
Horn       & 2.237 & 1.14 & 2.324 & 2.31 & 2.484 & 4.46 & 2.484 & 4.47 \\
Vivaldi    & 2.143 & 0.96 & 2.244 & 2.32 & -     & -    & 2.369 & 4.01 \\
Quasi-Yagi & 2.141 & 0.94 & 2.243 & 2.31 & 2.385 & 4.23 & 2.352 & 3.78 \\ \bottomrule
\end{tabular}%
}
\vspace{-0.4cm}
\end{table*}
\begin{figure*}
\centering
\subfloat[]{\includegraphics[width=0.25\textwidth]{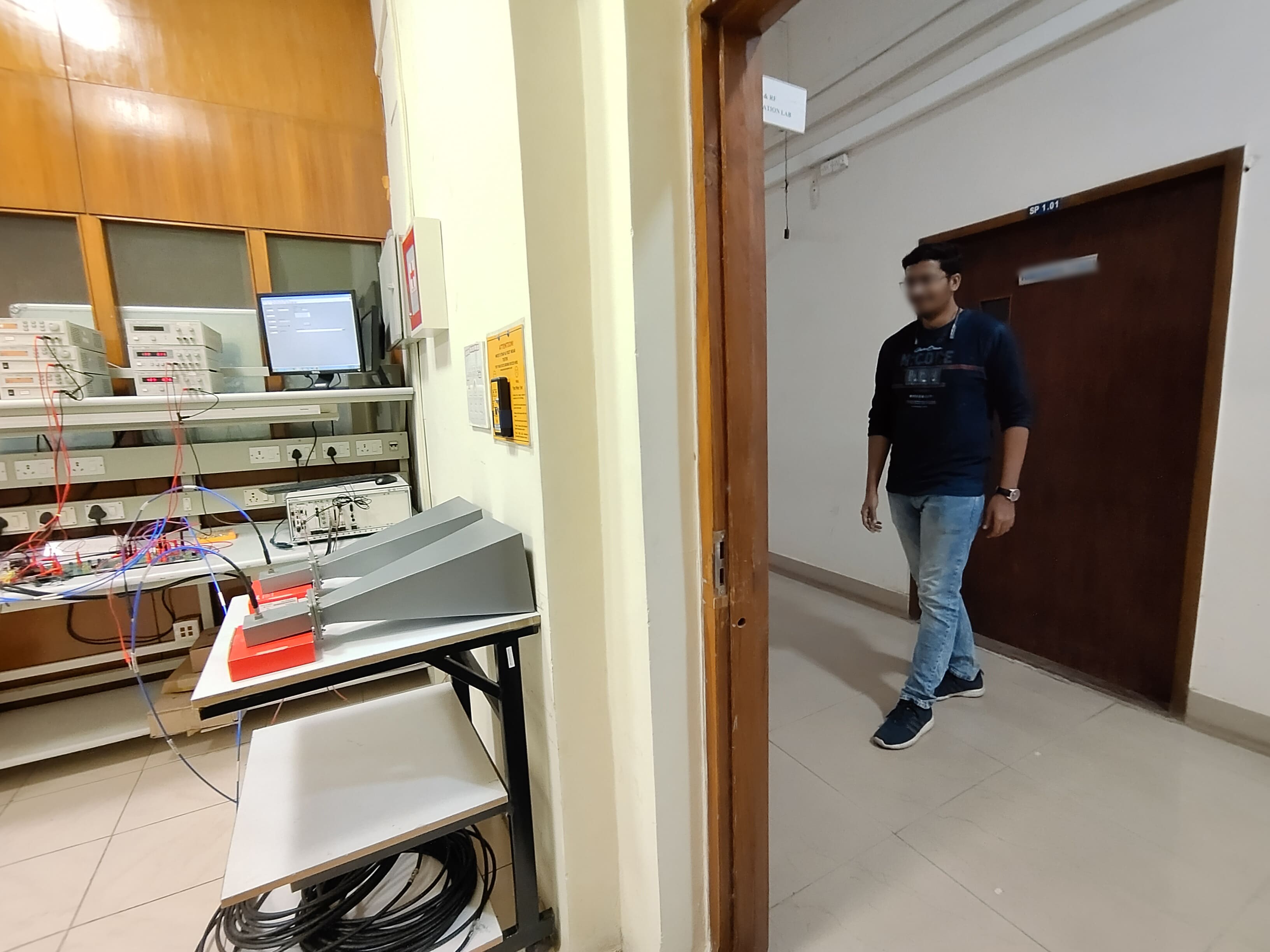}%
\label{4a}}
\hfil
\subfloat[]{\includegraphics[width=0.25\textwidth]{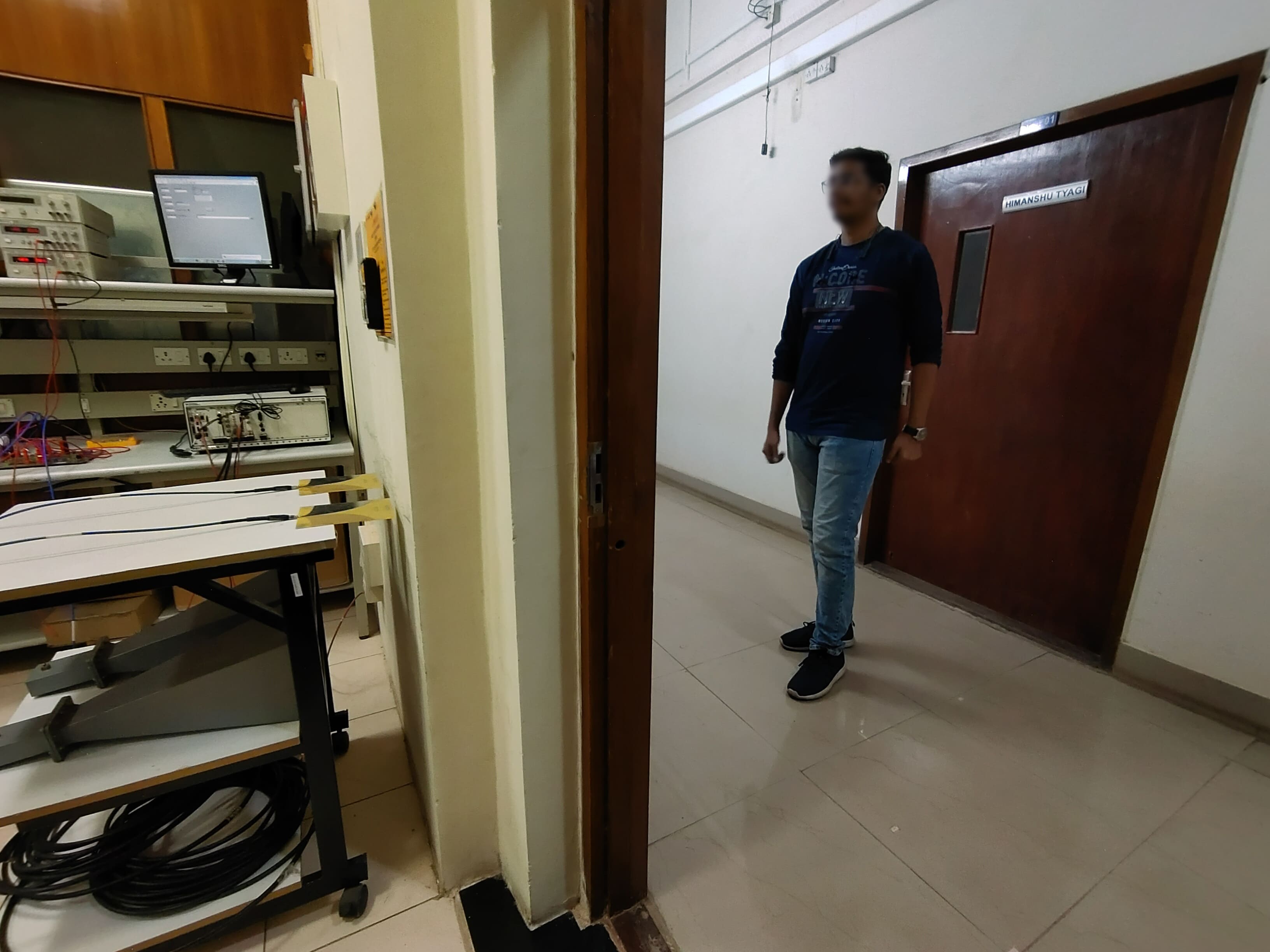}%
\label{4b}}
\hfil
\subfloat[]{\includegraphics[width=0.25\textwidth]{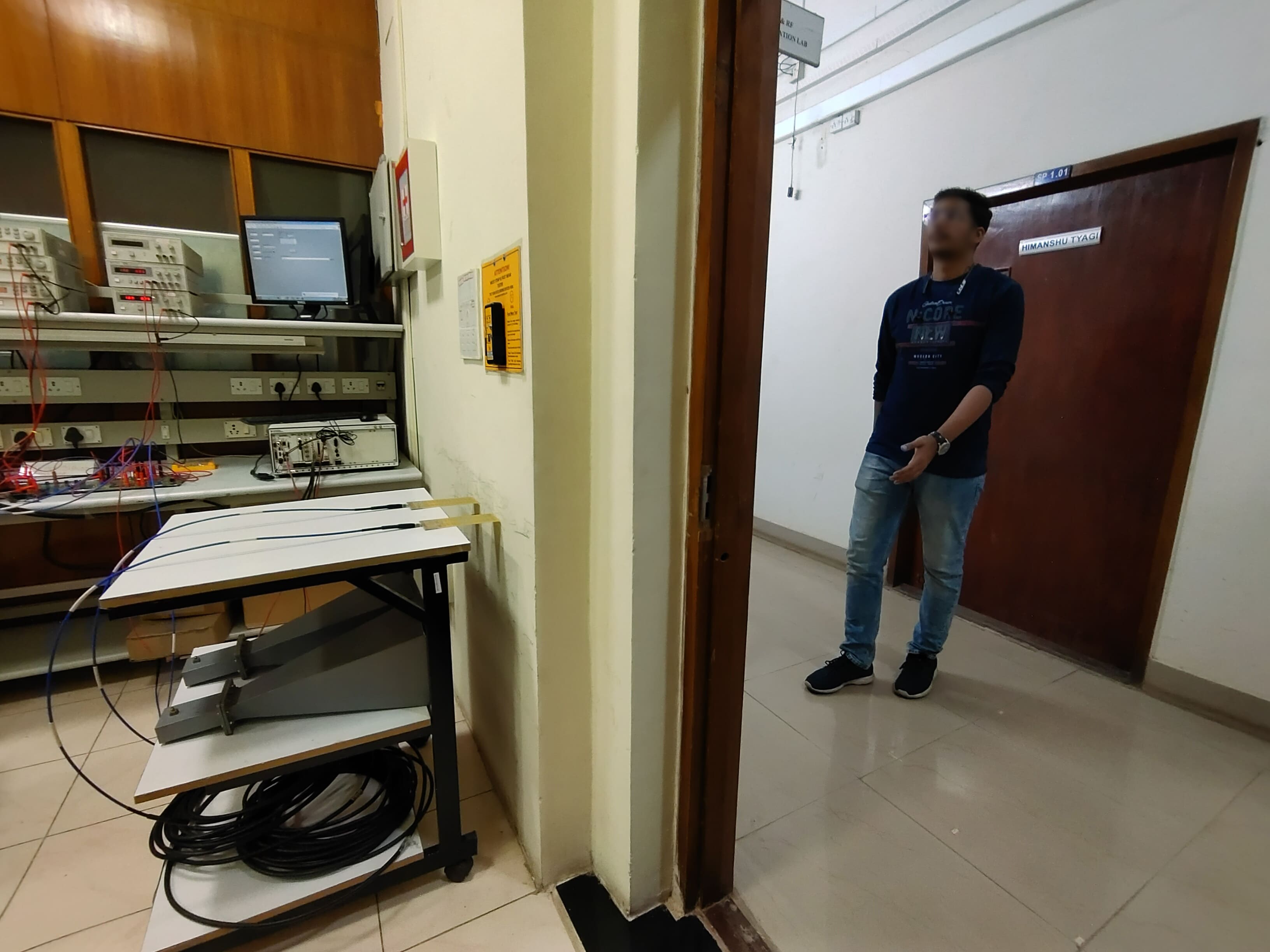}%
\label{4c}} \vspace{-0.4cm}
\hfil
\subfloat[]{\includegraphics[width=0.32\textwidth]{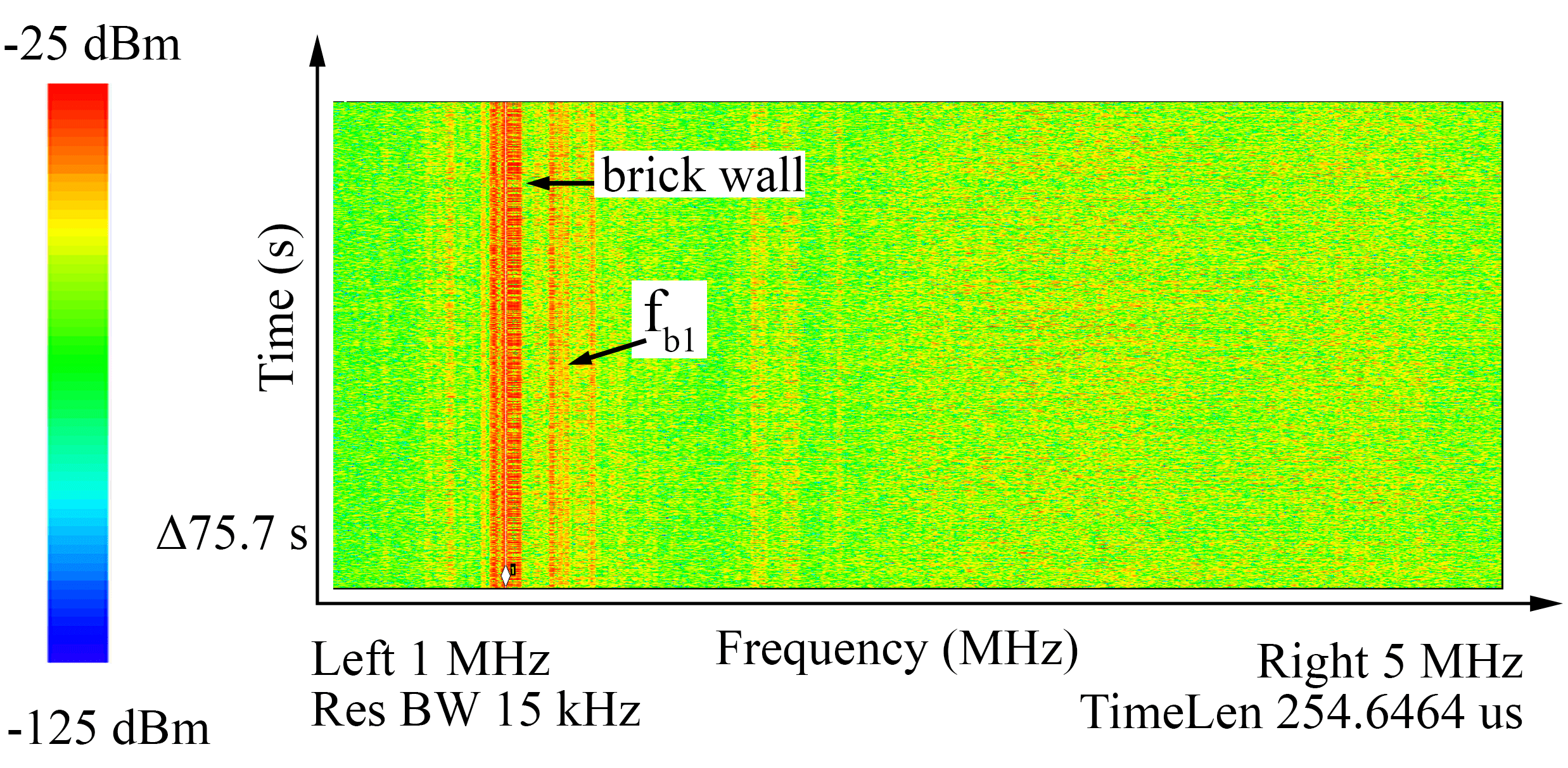}%
\label{4d}}
\hfil
\subfloat[]{\includegraphics[width=0.32\textwidth]{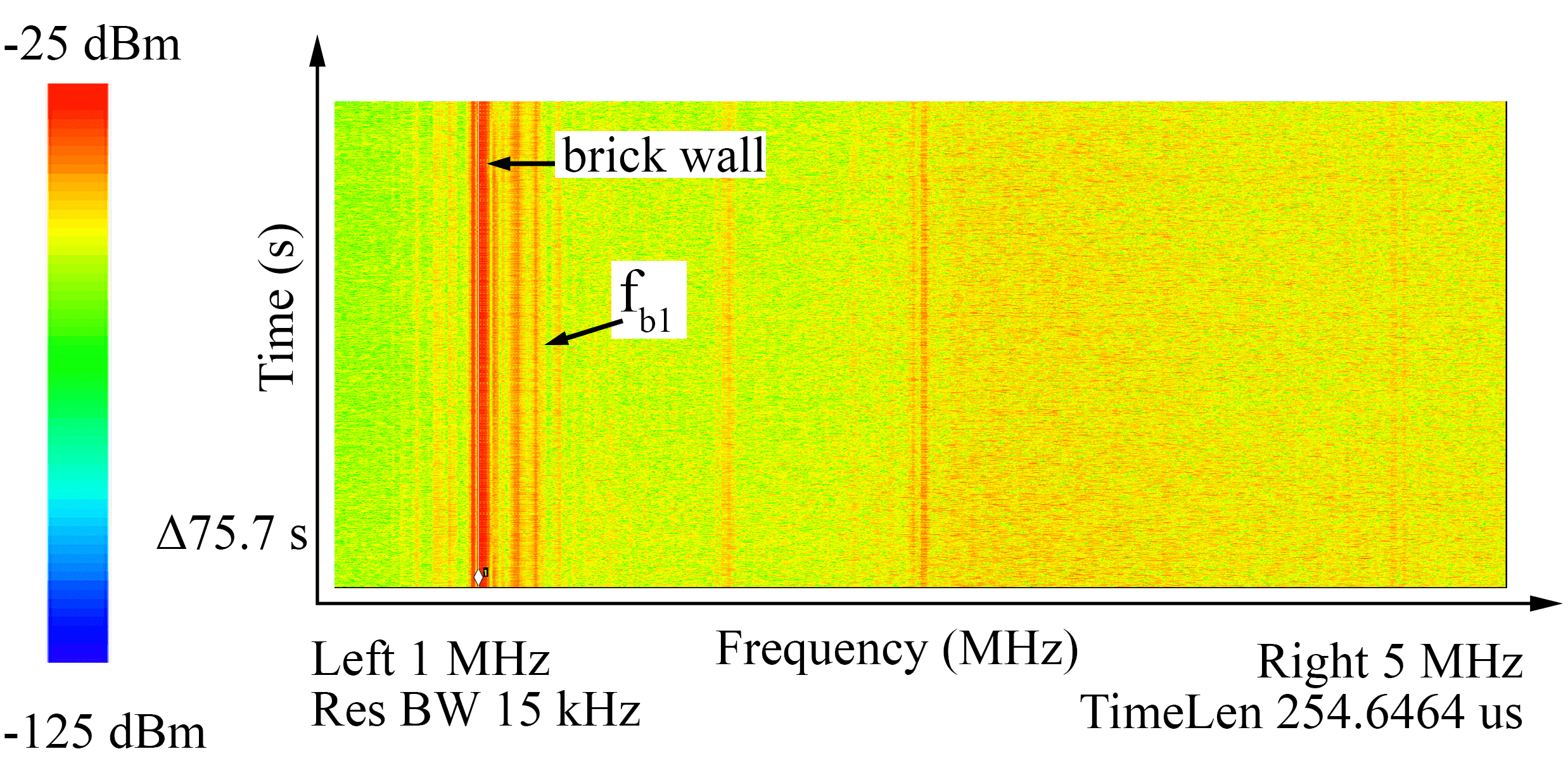}%
\label{4e}}
\hfil
\subfloat[]{\includegraphics[width=0.32\textwidth]{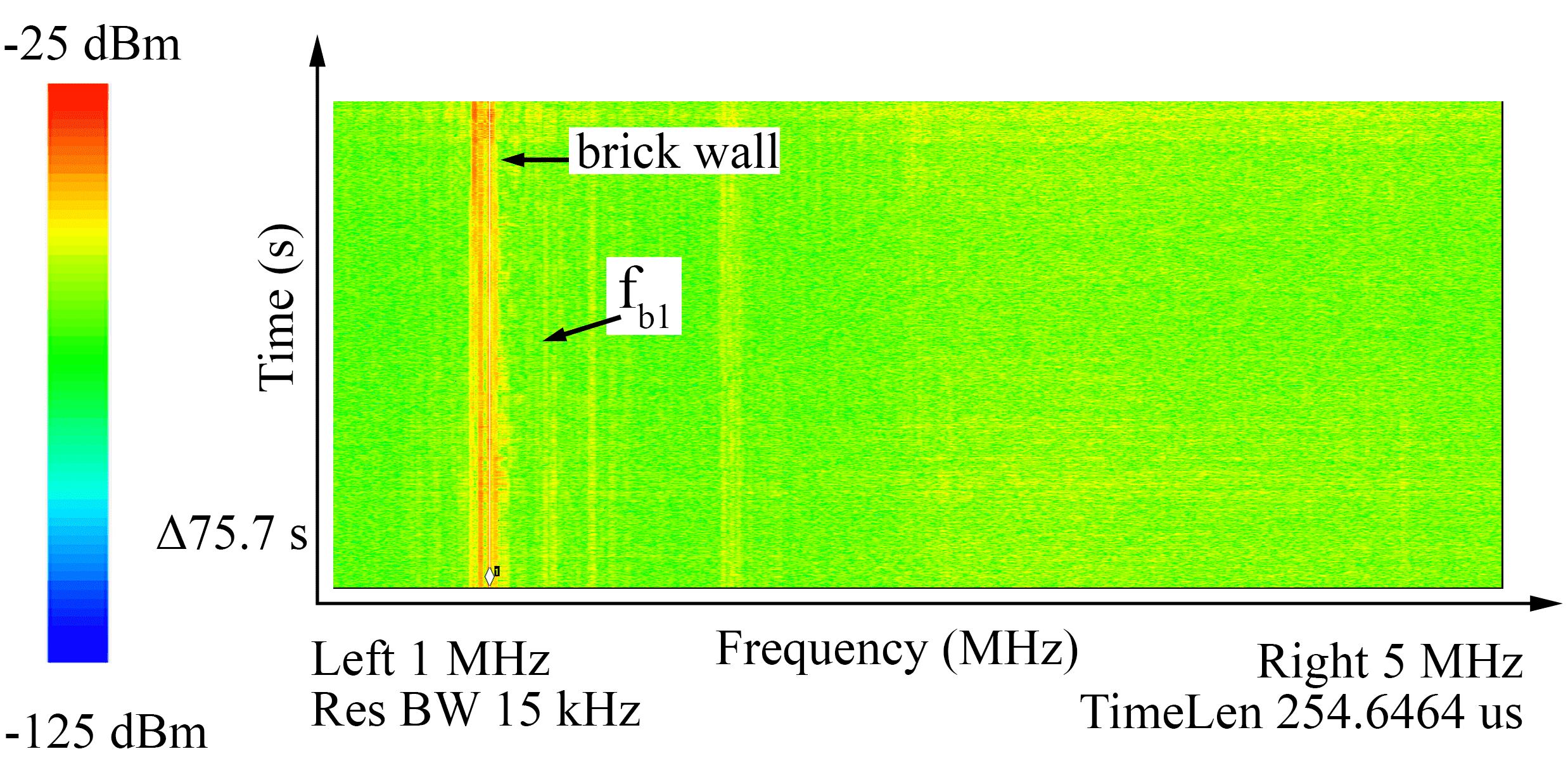}%
\label{4f}} \vspace{-0.4cm}
\hfil
\subfloat[]{\includegraphics[width=0.32\textwidth]{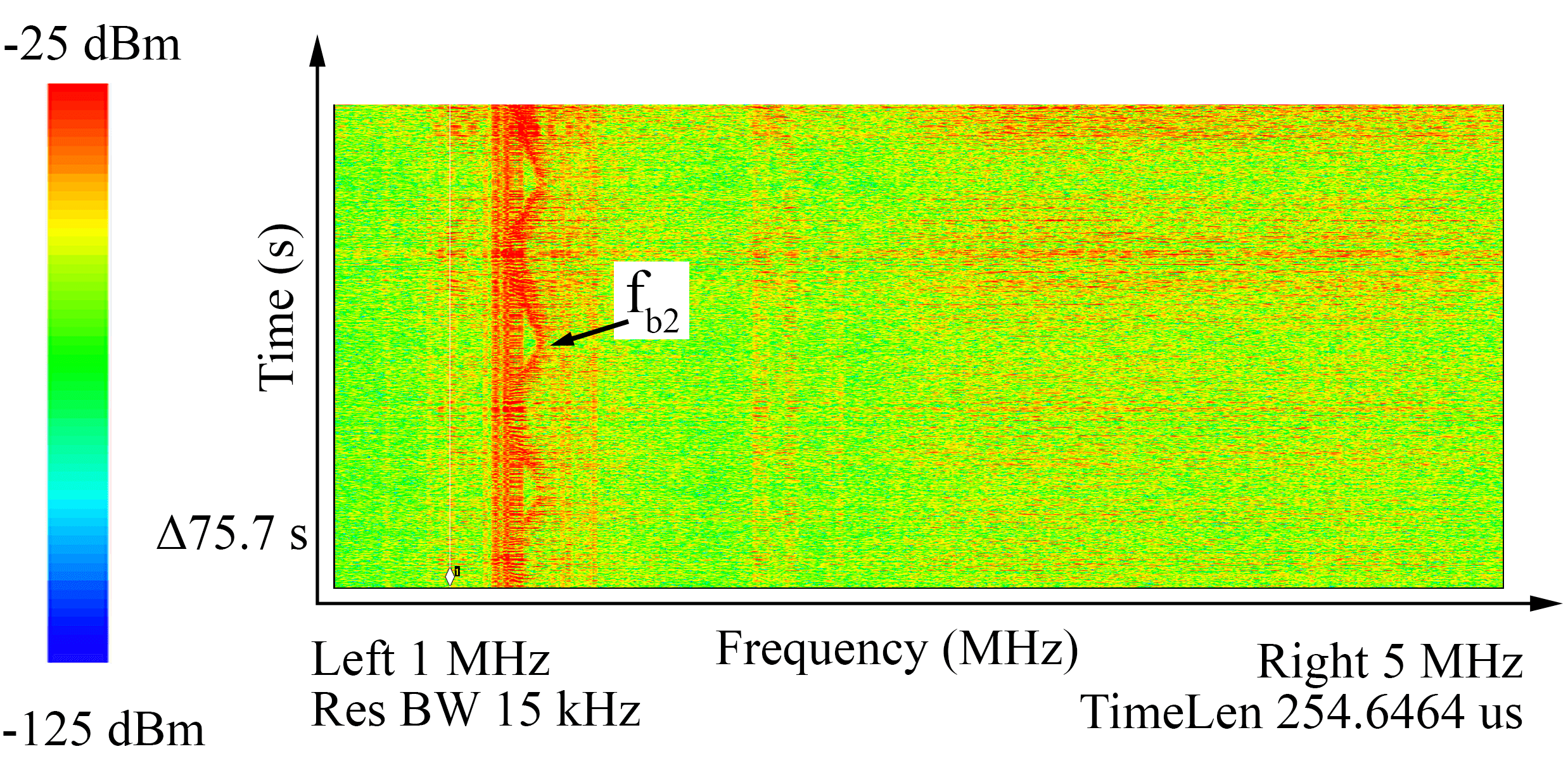}%
\label{4g}}
\hfil
\subfloat[]{\includegraphics[width=0.32\textwidth]{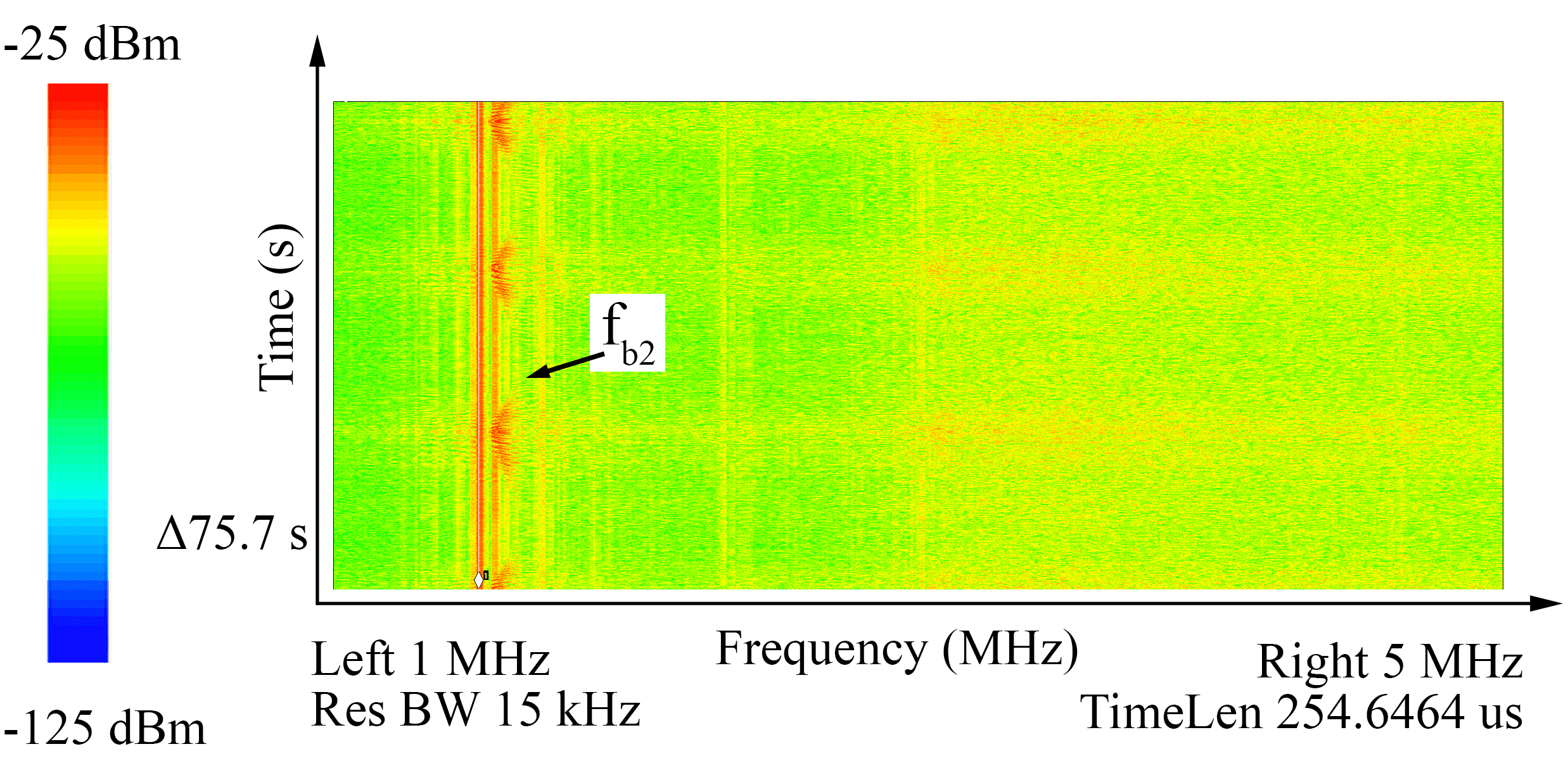}%
\label{4h}}
\hfil
\subfloat[]{\includegraphics[width=0.32\textwidth]{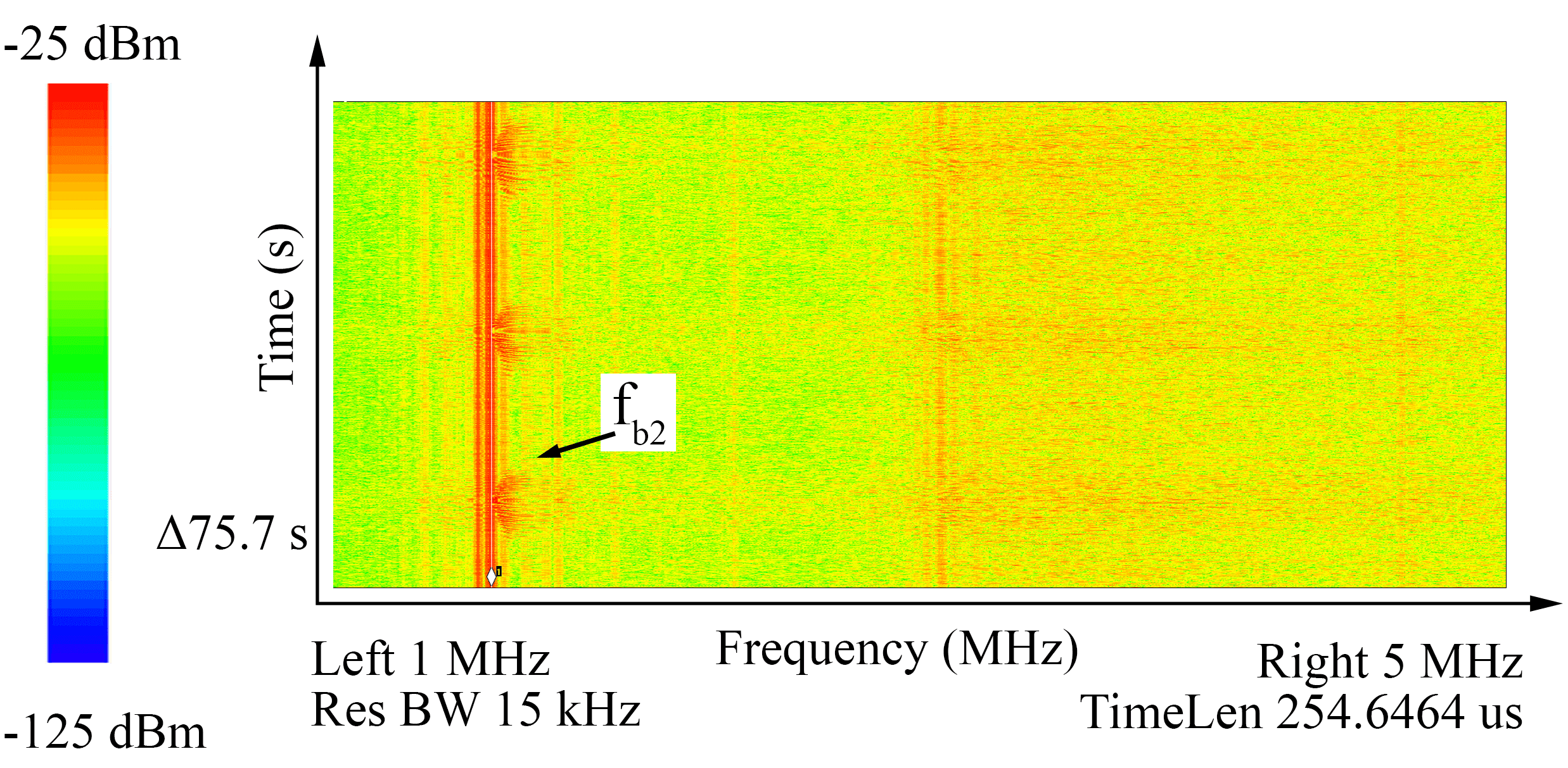}%
\label{4i}}
	\caption{Snapshots showing the experimental setup (a), (b), (c), spectrograms of the static environment (d), (e), (f) and spectrograms of the walking human (g), (h), (i) for thru-wall measurements for the horn, Vivaldi and quasi-Yagi antennas respectively.}
	\label{TWR} \vspace{-0.4cm}
\end{figure*}

\begin{table}
\centering
\caption{\textsc{Through wall Measurements}}
\label{tab: TWR}
\resizebox{0.3\columnwidth}{!}{%
\begin{tabular}{@{}ccccc@{}}
\toprule
\textbf{}  & \multicolumn{2}{c}{\textbf{Static}} & \multicolumn{2}{c}{\textbf{Person Walking}} \\ \midrule
\textbf{Antenna} &
  \textbf{\begin{tabular}[c]{@{}c@{}}$f_{b_1}$  \\ (MHz)\end{tabular}} &
  \textbf{\begin{tabular}[c]{@{}c@{}}$d_{obs_1}$ \\ (m)\end{tabular}} &
  \textbf{\begin{tabular}[c]{@{}c@{}}$f_{b_2}$\\ (MHz)\end{tabular}} &
  \textbf{\begin{tabular}[c]{@{}c@{}}$d_{obs_2}$ \\ (m)\end{tabular}} \\ \midrule
Horn       & 2.276            & 2            & 2.268                & 1.98                \\
Vivaldi    & 2.262            & 2.01            & 2.20                 & 1.2                \\
Quasi-Yagi & 2.254            & 2.01            & 1.03                 & 1.184                \\ \bottomrule
\end{tabular}%
}
\vspace{-0.6cm}
\end{table}
\begin{table}
\centering
\caption{\textsc{Comparison of antenna size}}
\label{tab:form factor}
\resizebox{0.3\columnwidth}{!}{%
\begin{tabular}{@{}ccccc@{}}
\toprule

\textbf{Antenna} &
  \textbf{\begin{tabular}[c]{@{}c@{}} Dimensions  \\ L$\times$W$\times$H (mm$^3$)\end{tabular}} &
  \textbf{\begin{tabular}[c]{@{}c@{}}Weight \\ (g)\end{tabular}} \\ \midrule
Horn       &  530$\times$240$\times$180          &     5000         \\
Vivaldi    & 180$\times$150$\times$1.6            &    79.9         \\
Quasi-Yagi & 180$\times$80$\times$1.6          &   42.6           \\ \bottomrule
\end{tabular}
}\vspace{-0.7cm}
\end{table}

\subsection{Through wall Measurements} \label{TWM}
For the through wall measurements, we imaged through a 40~cm thick outer wall (brick and mortar) of our laboratory, which faces another wall on the other side of a 2~m wide corridor (Fig. \ref{TWR}). 
We used a COTS power amplifier (MAAL-010200-TR3000), which provides a power gain of 10-20 dB in the frequency band of 2-2.6 GHz. The thru-wall-radar setup is shown in Figs. \ref{TWR}~(a), (b) and (c). 
Figs. \ref{TWR}~(d), (e) and (f) shows the snapshots of the spectrograms for the static environment. 
We can observe multiple closely spaced lines at a distance corresponding to that of the brick wall, as well as a straight line for the outer wall of the corridor at a distance of $d'$ = 2~m. 
Table \ref{tab: TWR} lists the observed beat frequency ($f_{b_1}$) and the corresponding distance of the back wall ($d_{obs_1}$) for all three antennas. 
The spectrogram signatures for the volunteer walking back and forth in front of of all the three antennas are shown in Figs. \ref{TWR}~(g), (h), and (i) respectively. 
Table \ref{tab: TWR} lists the maximum beat frequency ($f_{b_2}$) and the corresponding distance ($d_{obs_2}$) at which the moving target is identified. 
The power level of the spectrogram is kept the same for all the antennas (-25 to -125 dBm) to show the system's sensitivity.
Clearly, the signatures for the case of horn antenna are more prominent than that of the Vivaldi and quasi-Yagi antennas.
However, the quasi-Yagi antenna is able to identify moving targets behind the wall to distances that are quite comparable with the horn.

Using the above experiments, we have identified the range of the static objects (wall, wooden partition, metallic reflector, etc.) very close to the measured distance, as reported in Table \ref{tab:Environment A}, \ref{tab:Environment B: Closed Door} and \ref{tab: TWR}. Slight mismatches are due to the multiple reflections from a number of other static objects and the multi-path propagation of the waves since the experiments are conducted in a real world situation, instead of an ideal environment like an anechoic chamber. We have successfully observed the signatures of a walking human in two different environments and behind a brick wall using all the three antennas. Table \ref{tab:form factor} compares the physical dimension and weight of the three antennas used in this work. The signatures obtained from the planar antennas (Vivaldi and quasi-Yagi) may look faint when compared to the horn antenna on the same power scale but stronger signatures can be obtained by separately adjusting the mapping of the power levels for individual antennas to their respective color maps. Thus, the designed quasi-Yagi antenna can do the job as well as the horn antenna while significantly improving the portability and form factor of the complete radar system.

\section{Conclusions} \label{conclusion}
A compact quasi-Yagi antenna for TWR applications is presented in this paper. The quasi-Yagi antenna employs a modification of the ground plane for miniaturisation resulting in a 36\% size reduction. The performance of the designed antennas has been characterized in an ideal anechoic chamber environment and a real world scenario using a radar-on-chip based system.The TWR system is then tested in different environments to study the environmental effects. The radar can successfully identify movement behind a wooden partition and a brick wall. The quasi-Yagi antenna, when compared to a horn antenna, exhibits similar levels of performance when used in a real world system. The proposed antenna is planar, light-weight and compact in size, when compared to a bulky horn antenna and thus helps in making the radar-on-chip system portable and easy to deploy.
\section*{Acknowledgement} \label{Acknowledgement}
The authors express their gratitude to Mr. Himanshu B. Sandhibigraha, Mr. Alok C. Joshi and Mr. Rituraj Kar for their help with the experiments.


\bibliography{wileyNJD_Doc}%







\end{document}